\documentclass[10pt,leqno]{amsart}
\usepackage{indentfirst,csquotes}
\usepackage{listings}
\usepackage{xcolor}
\usepackage{subcaption}
\usepackage{adjustbox}
\usepackage{multirow}
\usepackage[margin=1in]{geometry}
\usepackage{hhline}
\usepackage{booktabs}
\usepackage{makecell}
\usepackage{chngcntr}
\usepackage{algorithm}
\usepackage[T1]{fontenc}
\usepackage{svg}
\usepackage{tikz}
\usetikzlibrary{patterns, decorations.pathmorphing}
\usepackage{pgfplots}
\usepackage{pgfplotstable}
\pgfplotsset{compat=1.18}
\usepgfplotslibrary{groupplots}
\usepackage{pgf-pie}
\usepackage{xcolor} 
\usetikzlibrary{fit}
\usetikzlibrary{backgrounds}
\definecolor{myorange}{RGB}{233,202,175}
\definecolor{mygreen}{RGB}{185,217,188}
\definecolor{myblue}{RGB}{174,224,228} 
\definecolor{myorange_dark}{RGB}{233,122,75}  
\definecolor{mygreen_dark}{RGB}{85,217,88}    
\definecolor{myblue_dark}{RGB}{74,124,228}    
\definecolor{myred_dark}{RGB}{139,0,0}
\definecolor{myred}{RGB}{217,85,98}
\definecolor{darkgreen}{RGB}{0, 100, 0}
\definecolor{TUMBlue1}{HTML}{000953}  
\definecolor{TUMBlue2}{HTML}{005199}  
\definecolor{TUMBlue3}{HTML}{0065BF}  
\definecolor{TUMBlue4}{HTML}{4283B0}  
\definecolor{TUMBlue5}{HTML}{85BAE0}  
\definecolor{TUMBlue6}{HTML}{B4DCFF}  

\usepackage{lipsum}
\usepackage{lineno}
\usepackage{amsfonts}
\usepackage{amsmath}
\usepackage{amssymb}
\usepackage{appendix}
\usepackage{graphicx}
\usepackage{epstopdf}
\usepackage{caption}
\usepackage{subcaption}
\usepackage{algorithm}
\usepackage[noend]{algpseudocode}
\usepackage{multirow}
\usepackage{booktabs}
\usepackage{hyperref}
\usetikzlibrary{shapes.geometric, arrows}
\usepackage{cleveref}

\tikzstyle{startstop} = [ellipse, minimum width=2cm, minimum height=1cm, text centered, draw=black, fill=TUMBlue3!70]
\tikzstyle{process} = [rectangle, text width=5.5em, text badly centered, minimum width=3cm, minimum height=1cm, text centered, draw=black, fill=TUMBlue6!80]
\tikzstyle{decision} =  [diamond, draw,  minimum width=7em,
    text width=4.5em, text badly centered, node distance=3cm, inner sep=0pt,fill=TUMBlue3!50]
\tikzstyle{arrow} = [thick,->,>=stealth]
\tikzstyle{processblue} = [rectangle, minimum width=2.5cm, minimum height=1cm, text centered, draw=black, fill=TUMBlue6!60]
\tikzstyle{line} = [draw, -latex']
\tikzstyle{cloud} = [draw, ellipse,fill=red!20, node distance=3cm,
    minimum height=2em]
\tikzstyle{subroutine} = [rectangle split, rectangle split horizontal,
    rectangle split parts=3, draw, 
    minimum width=6em, 
    minimum height=1.5cm, text centered, outer sep=0, fill=TUMBlue4!80]

\newcommand{\R}{\ensuremath{\mathbb{R}}}

\newcommand{\K}{\ensuremath{\mathbf{K}}}
\newcommand{\C}{\ensuremath{\mathbf{C}}}
\newcommand{\M}{\ensuremath{\mathbf{M}}}
\newcommand{\F}{\ensuremath{\mathbf{f}}}
\newcommand{\G}{\ensuremath{\mathbf{g}}}
\newcommand{\V}{\ensuremath{\mathbf{V}}}
\newcommand{\p}{\ensuremath{\mathbf{p}}}
\newcommand{\x}{\ensuremath{\mathbf{x}}}

\newcommand{\xr}{\ensuremath{\mathbf{x}_r}}

\newcommand{\Krk}{\ensuremath{\mathbf{K}_{r,k}}}
\newcommand{\Crk}{\ensuremath{\mathbf{C}_{r,k}}}
\newcommand{\Mrk}{\ensuremath{\mathbf{M}_{r,k}}}
\newcommand{\Frk}{\ensuremath{\mathbf{f}_{r,k}}}
\newcommand{\Grk}{\ensuremath{\mathbf{g}_{r,k}}}

\newcommand{\pk}{\ensuremath{\mathbf{p}_{k}}}
\newcommand{\Vk}{\ensuremath{\mathbf{V}_{k}}}

\begin{document}
\title[Efficient Optimization of Matrix-Interpolatory Reduced-Order Models]{Uncertainty-Aware Calculation of Analytical Gradients of Matrix-Interpolatory Reduced-Order Models for Efficient Structural Optimization
} 

\author{Marcel Warzecha$^1$, Sebastian Resch-Schopper$^{2}$, and Gerhard Müller$^2$}

\date{\today \\[2mm]
    $^1$Technical University of Munich, School of Engineering and Design, Professorship for Data-driven Materials Modeling, Boltzmannstr. 15, 85748 Garching b. München (marcel.warzecha@tum.de) \\%
    $^2$Technical University of Munich, School of Engineering and Design, Chair of Structural Mechanics, Arcisstr. 21, 80333 München (sebastian.resch-schopper@tum.de, gerhard.mueller@tum.de)\\[2mm]
    Corresponding author: Marcel Warzecha (marcel.warzecha@tum.de)
}

\begin{abstract}
This paper presents an adaptive sampling algorithm tailored for the optimization of parametrized dynamical systems using projection-based model order reduction. Unlike classical sampling strategies, this framework does not aim for a small approximation error in the global sense but focuses on identifying and refining promising regions early on while reducing expensive full order model evaluations. The algorithm is tested on two models: a Timoshenko beam and a Kelvin cell, which ought to be optimized in terms of the system output in the frequency domain. For that, different norms of the transfer function are used as the objective function, while up to two geometrical parameters form the vector of design variables. The sampled full order models are reduced using the iterative rational Krylov algorithm and reprojected into a global basis. Subsequently, the models are parametrized by performing sparse Bayesian regression on matrix entry level of the reduced operators. Thompson sampling is carried out using the posterior distribution of the polynomial coefficients in order to account for uncertainties in the trained regression models. The strategy deployed for sample acquisition incorporates a gradient-based search on the parametrized reduced order model, which involves analytical gradients obtained via adjoint sensitivity analysis. By adding the found optimum to the sample set, the sample set is iteratively refined. Results demonstrate robust convergence towards the global optimum but highlight the computational cost introduced by the gradient-based optimization. The probabilistic extensions seamlessly integrate into existing matrix-interpolatory reduction frameworks and enable the analytical calculation of gradients under uncertainty.
\end{abstract}

\maketitle

\let\thefootnote\relax




\section{Introduction}%
\label{sec:intro}

The optimization of technical systems is of interest in numerous engineering disciplines. The behavior of these systems is thereby often evaluated using numerical approaches such as the Finite Element Method (FEM). For dynamical systems, this simulation is often computationally demanding, as it requires repeatedly solving large systems in either the time or frequency domain. When performing an optimization, the computational cost increases significantly because simulations must be conducted for numerous parameter variations. This motivates the use of techniques that reduce the associated computational burden. \\
Projection-based model order reduction (MOR) provides one such acceleration strategy by approximating the high-fidelity solution in a low-dimensional subspace. The full-order operators are projected onto a reduced basis, resulting in a small system that can be solved efficiently while still accurately capturing the behavior of the full-order model (FOM) \cite{Benner2015}. \\
For parameter-dependent problems, the reduced-order model should also contain the parametric dependencies, which can be achieved by using parametric MOR (pMOR) methods. Existing approaches can be broadly divided into global and local strategies. Global methods rely on a single reduced basis that is used in the entire parameter domain. For this method to be efficient, the parametric dependence should admit an efficient affine representation, i.e., a linear combination of constant matrices with scalar functions of the parameters. This can often be achieved easily for material parameters, but might be hard or impossible to obtain for geometric parameters. In the absence of an efficient affine decomposition, hyper-reduction methods such as the discrete empirical interpolation method \cite{Chaturantabut2010} or the energy-conserving sampling and weighting method \cite{Farhat2014} can be used to accelerate the evaluation of the reduced model. Local pMOR techniques instead do not rely on a global basis but interpolate parameter-dependent reduced quantities such as the reduced bases \cite{Amsallem2008, Goutaudier2023}, the reduced system matrices \cite{Amsallem2009, Panzer2010, Amsallem2011}, or the (reduced) transfer functions \cite{Baur2009,Ionita2014, Rodriguez2023}. \\
This work focuses on a mixture between global pMOR and matrix-interpolation-based pMOR as originally proposed in \cite{Degroote2010}: First, full-order models and corresponding reduced bases are computed for a set of sample points. Afterwards, one global basis is computed onto which all sampled FOMs are projected. The resulting reduced system matrices are then interpolated to efficiently predict the ROM for a queried parameter point. The advantage of interpolating the reduced system matrices is that it avoids the need for an affine representation of the parametric dependency. Furthermore, interpolating the reduced system matrices yields an explicit expression for the predicted reduced system matrices in terms of the parameters. This can be used to analytically compute the gradients of the reduced system matrices and, consequently, the gradient of the objective function \cite{Choi2020}. Using a global basis for the reduction of the sampled FOMs avoids the need to transform the individually reduced systems to a common coordinate system, which is otherwise necessary for a meaningful interpolation \cite{Amsallem2009, Panzer2010, Amsallem2011}. For large parameter ranges, a transformation to the same coordinate system might be impossible, which then introduces inconsistencies in the training data for the matrix interpolation \cite{Resch-Schopper2026}. \\
One important aspect of pMOR approaches that determines the accuracy of the resulting parametric reduced-order model (pROM) is the distribution of the sample points. For global pMOR, a greedy approach that places the samples at the points where the estimated error is highest is often adopted \cite{Prud'homme2002,Bui-Tanh2008}. In the context of pMOR by matrix interpolation, both one-shot and adaptive sampling methods have been proposed. Regarding the former, sparse grids \cite{Geuss2014,Geuss2015,Brunsch2017} and sample distributions based on numerical integration schemes \cite{Mencik2021,Mencik2024} have proven beneficial. For adaptive sampling strategies, one criterion that can be used is the principal angles between the subspaces spanned by the sampled reduced bases \cite{Bazaz2015,Varona2017,Resch-Schopper2026}. Similar to greedy sampling strategies in global pMOR, approaches based on error estimators have been proposed in \cite{Liu2019_sampling,Liu2019_samplingCMS,Choi2020,Goizueta2021,Goizueta2022}. All of these approaches aim to distribute the samples such that the resulting pROM is accurate across the entire parameter space. \\
In the context of an optimization, however, one might not require a high accuracy of the pROM in the entire parameter space, but only around the optimum. Specifically, generating the samples and building the final pROM might already take more computational effort than performing the optimization with the FOM or individual non-parametric ROMs. A computationally more efficient framework would therefore be achieved when the adaptive sampling is driven by the optimization. Such an approach has been proposed in \cite{Leite2019}. There, the initial pROM is generated from a small set of samples that form a simplex. This pROM is then used for the optimization, and if the simplex is left during the optimization, a new sample point is added so that a new simplex in the region of the intermediate optimum can be constructed. This way, new samples are only placed around the optimization trajectory. This approach, however, focuses solely on exploitation, so improving the currently best solution, but not on exploration, which refers to searching for the optimum in other regions. \\
We therefore propose a novel framework for optimization-based adaptive sampling that balances exploration and exploitation. The central observation underlying this work is that a unifying methodology of uncertainty-aware exploration paired with gradient-consistent exploitation has not been proposed yet. On a high-level, the methodology can be described as follows: FOMs are sampled selectively and reduced using the Iterative Rational Krylov Algorithm (IRKA) \cite{Gugercin2008,Wyatt2012}. The obtained reduced operators are reprojected into a global basis and used as training data for a sparse Bayesian regression (SBR) \cite{tipping2001} subroutine. The resulting probabilistic models are then utilized to drive the sample acquisition for extending the set of existing FOMs. Candidate points are identified by solving a gradient-based optimization problem on the pROM level, where analytical gradients are obtained using adjoint sensitivity analysis. By performing a Thompson sampling (TS)  \cite{Thompson1933} on the posterior distribution of the SBR routine, the uncertainty of the interpolated models is accounted for in the subsequent gradient calculation. Acquired samples are incorporated iteratively and refine the overall pROM in regions relevant for the optimization while accounting for the uncertainty of the regression models. By deploying this strategy, this approach intentionally sacrifices a globally accurate pROM in favor of a fast convergence towards promising regions.\\
The remainder of this paper is organized as follows. Section~\ref{sec:theory} focuses on the problem definition by introducing the underlying PDE describing a dynamical system as well as the objective function that is defined in the frequency space. Sections~\ref{sec:pMOR} and~\ref{sec:GB-Optimization} present the relevant theory in MOR and gradient-based optimization problems. In Section~\ref{sec:methodology}, a detailed description of the deployed methodology is provided, while numerical examples and results are discussed in Section~\ref{sec:Results}. Finally, conclusions are drawn and perspectives for future work are given in Section~\ref{sec:conclusions}.

\section{Theoretical Background}%
\label{sec:theory}
\subsection{Problem Definition}
\label{sec:ProblemDefinition}

We consider the optimization of parameter-dependent, linear time-invariant dynamical systems with single input and single output (SISO). These systems are represented in second-order form in the frequency domain as follows: 
\begin{equation}
   \Sigma(\p): \begin{cases}
		s^2\M(\p) \x(\p,s) + s\C(\p) \x(\p,s) + \K(\p) \x(\p,s) &= \F(\p) u(s), \\
		\hfill y(\p,s) &= \G(\p)\mathbf{x}(\p,s),
		\end{cases} \label{eq:FOM}
\end{equation}
The matrices $\mathbf{M}, \, \mathbf{C}, \, \mathbf{K}: \Omega \rightarrow \mathbb{R}^{n \times n}$ are the parameter-dependent mass, damping and stiffness matrices, which depend on $d$ parameters $\p = [p_1, p_2, \dots , p_d] \in \Omega \subset \R^d$, where $\Omega$ denotes a bounded domain. The vector of degrees of freedom is given by $\x \in \mathbb{R}^n$, $n$ denotes the number of degrees of freedom, $s \in \mathbb{C}$ is the complex frequency, $u(s) \in \R$ and $\F: \R^d \rightarrow \R^{n \times 1}$ the input and $y(\mathbf{p},s) \in \C$ and $\G: \R^d \rightarrow \R^{1 \times n}$ the output.  For the sake of simplicity, we restrict ourselves to single-input single-output (SISO) systems, but the concepts presented in the following may be applied to multiple-input multiple-output (MIMO) systems as well. In problems of structural dynamics, the input and the output vector are usually parameter-independent, so their dependency on $\p$ will be dropped in the following. \\
A common objective of an optimization of these systems is to minimize the output in a specific frequency range $s_{opt}$ \cite{Venini2016}. For a discrete set of $n_s$ frequency points in that range, we thus define the objective function $R(\p)$ as follows, using the generalized $L_k$-norm of the output:
\begin{equation}
    R(\p) = \sqrt[k]{\sum_{i}^{n_s}  (|\mathbf{g}\mathbf{x}(\p,s_{i})|)^{k}}.
    \label{eq:objective_k}
\end{equation}

\section{Parametric Model Order Reduction}
\label{sec:pMOR}

The system shown in \Cref{eq:FOM} can comprise a large number of degrees of freedom for complex problems, making it computationally very expensive to solve for multiple instances in frequency. In the context of an optimization, the system furthermore needs to be solved for various configurations of the parameters, which can become infeasible. Methods to reduce the computational effort required for these systems are therefore needed to enable such multi-query applications.

\subsection{Projection-based Model Order Reduction}
\label{sec:MOR}

A state-of-the-art method to reduce the computational effort of the full-order system shown in \Cref{eq:FOM} for a specific parameter point $\p_k$ is projection-based model order reduction (MOR). The idea of these approaches is to find a lower-dimensional subspace spanned by a reduced basis $\Vk \in \mathbb{R}^{n \times r}$ in which the full solution can be approximated well. This way, the size of the problem can be reduced to $r \ll n$. The reduced system is then obtained by projecting the full-order matrices onto the reduced basis:
\begin{equation}
\begin{aligned}
    \Mrk &= \Vk^\intercal \M(\pk) \Vk, \qquad &\Crk &= \Vk^\intercal \C(\pk) \Vk, \qquad \Krk = \Vk^\intercal \K(\pk) \Vk \\
    \Frk &= \Vk^\intercal \F(\pk), \qquad &\Grk &= \G(\pk) \Vk.
    \label{eq:Projection}
\end{aligned}
\end{equation}
The reduced system then takes the same form as the full-order system:
\begin{equation}
    \Sigma_r(\pk): \begin{cases}
		s^2\Mrk \x_r(s) + s\Crk \x_r(s) + \Krk \xr(s) &= \Frk u(s), \\
		\hfill y_r(s) &= \Grk \xr(s).
		\end{cases} \label{eq:ReducedSystem}
\end{equation}
The accuracy of the reduced-order model strongly depends on the choice of the reduced basis. Over the past decades, various methods for generating the reduced basis have been developed, such as modal methods \cite{Tiso2021}, moment matching \cite{Benner2021a}, or Proper Orthogonal Decomposition \cite{Antoulas2005}. 
In moment matching, the reduced basis is computed such that the value and the derivative up to order $n_o$ of the reduced and the full transfer function match for a set of expansion frequencies $s_1, \dots, s_r$, i.e.
\begin{equation}
    \frac{\mathrm{d}^k}{\mathrm{d}s^k} H(s_i) = \frac{\mathrm{d}^k}{\mathrm{d}s^k} H_r(s_i) \quad \text{for} \quad k = 0, 1, \ldots, n_o.
\end{equation} 
The choice of the expansion frequencies has a major impact on the accuracy of the resulting reduced-order model. One approach to obtain these is by using the iterative rational Krylov algorithm (IRKA), which aims at finding optimal expansion frequencies in an $H_2$-sense. It has been initially proposed for first-order systems \cite{Gugercin2008} and was later extended to second-order systems \cite{Wyatt2012}. The algorithm starts with a set of $r$ arbitrarily chosen expansion frequencies closed under complex conjugation. The reduced basis is then computed from the full solution at these expansion frequencies:
\begin{equation}
    \left( s_i^2 \M + s_i \C + \K \right)^{-1} \F \in \mathrm{Ran}(\V). \label{eq:BasisIRKA}
\end{equation}
Next, this basis is used to reduce the full system as shown in \Cref{eq:Projection}. To obtain the expansion frequencies for the next iteration, the eigenvalues of the reduced system are computed, which results in $2r$ values. In order not to increase the size of the reduced model in each iteration, the mirror images of only $r$ of these values are chosen as new expansion frequencies. In this work, the mirror images of the eigenvalues closest to the imaginary axis are chosen so that the reduced-order model is most accurate for low frequencies. However, the expansion frequencies could also be chosen in a specific targeted frequency range \cite{Aumann2022}. Afterwards, a new basis is computed again using \Cref{eq:BasisIRKA} and the steps are repeated until the expansion frequencies converge.

\subsection{Parametric Model Order Reduction by Matrix Interpolation}
\label{sec:pMOR_MI}

To incorporate parametric dependencies in the reduced-order model, parametric model order reduction (pMOR) methods are required. These can be distinguished into global and local methods. In the former, one global reduced basis is computed that is used to project all full-order systems in the complete parameter space, whereas local methods interpolate some reduced quantities, such as the reduced basis, the reduced operators, or the reduced transfer function. In this work, we combine global pMOR with local pMOR by matrix interpolation \cite{Amsallem2011, Panzer2010} as initially proposed in \cite{Degroote2010}: For a set of samples in the parameter space, we individually compute a reduced basis for each sample and concatenate them to a global reduced basis. Then, all sampled full-order models are projected onto this global basis. Finally, the entries of these reduced operators are interpolated so that a reduced system can be predicted efficiently in the online phase. This way, no affine representation of the parametric dependency of the full operators is required for the method to be efficient, as is usually the case for global pMOR methods. Furthermore, this approach leads to an explicit representation of the parametric dependency of the reduced operators, which will be made use of later. Compared to pMOR by matrix interpolation, the advantage of this approach is that the sampled reduced operators all lie in the same coordinate system, as they are projected onto the same global basis. This way, a congruence transformation of the individually sampled reduced basis, which is otherwise necessary in these approaches, is avoided.

\subsubsection{Sparse Bayesian Regression}
\label{sec:sbr}
The entries of the reduced operators can be interpolated with any interpolation or regression method. In this work, we are using sparse Bayesian regression (SBR) as provided in \cite{tipping2001} for two reasons: Firstly, the basis of this approach is linear regression, which allows an approximation of the parametric dependency of the reduced operators with simple functions such as polynomials. This will be helpful later on for computing derivatives of the objective function efficiently. Secondly, the coefficients of the regression model are treated as uncertain parameters in SBR, which allows to account for the uncertainty of the predicted reduced operators. In combination with an acquisition function, this enables balancing exploration and exploitation in the active learning strategy that we propose. \\
In linear regression, an entry $a$ of one of the reduced operators is approximated by 
\begin{equation}
    a(\p) = \sum_{i=1}^{n_f} f_i(\p) \beta_i + \varepsilon, \label{eq:Regression}
\end{equation}
where $f_i$ are scalar functions that can be chosen arbitrarily, $\beta_i$ are the coefficients we want to find so that the prediction matches the sampled values well, and $\varepsilon$ is the error made in the prediction. In standard linear regression, the coefficients $\beta$ are computed by solving the least-squares problem
\begin{equation}
    \mathbf{X}^\intercal \mathbf{X} \pmb{\beta} = \mathbf{X}^\intercal \mathbf{a},
\end{equation}
where $\mathbf{X}$ contains the evaluation of the scalar functions $f_i$ at the sample points $p_k$:
\begin{equation}
    \mathbf{X} = \begin{bmatrix} f_1(\p_1) & f_2(\p_1) & \cdots & f_{n_f}(\p_1) \\
    f_1(\p_2) & f_2(\p_2) & \cdots & f_{n_f}(\p_2) \\
    \vdots & \vdots & \ddots & \vdots \\
    f_1(\p_K) & f_2(\p_K) & \cdots & f_{n_f}(\p_K) \end{bmatrix}, \quad \pmb{\beta} = \begin{bmatrix} \beta_1 \\ \beta_2 \\ \vdots \\ \beta_{n_f} \end{bmatrix}, \quad \text{ and } \quad \mathbf{a} = \begin{bmatrix} a_1 \\ a_2 \\ \vdots \\ a_K \end{bmatrix}.
\end{equation}     
To introduce uncertainties, we define a probabilistic model for $\varepsilon$ assuming that it is Gaussian distributed with zero mean and standard deviation $\sigma$, so
\begin{equation}
    P(\pmb{\varepsilon}\mid \sigma^2) = \mathcal{N}(0; \sigma^2).
    \label{eqn:gaussian_distr}
\end{equation}
Applying Bayes' Theorem to the regression problem results in 
\begin{equation}
    P(\pmb{\beta}\mid \mathbf{a}) = \frac{P(\mathbf{a}\mid \mathbf{X},\pmb{\beta},\sigma^2)\, P(\pmb{\beta})}{P(\mathbf{a})}.
    \label{eqn:bayes_theorem_reg}
\end{equation}
The coefficients $\pmb{\beta}$ are now found by maximizing the posterior $P(\pmb{\beta}\mid \mathbf{a})$, which results in the maximum a posteriori (MAP) estimate. Since $\pmb{\beta}$ only influences the terms in the nominator in \Cref{eqn:bayes_theorem_reg}, this can be written as
\begin{equation}
    \pmb{\beta}^* = \pmb{\beta}_{(\text{MAP})} =  \underset{\pmb{\beta}}{\arg \max} \, P(\pmb{\beta}\mid  \mathbf{y}) = \underset{\pmb{\beta}}{\arg \max} \, \left[P(\mathbf{y}\mid \mathbf{X},\pmb{\beta},\sigma^2)\, P(\pmb{\beta})\right]. 
    \label{eqn:beta_MAP}
\end{equation}
To solve this optimization, a prior distribution $P(\pmb{\beta})$ is required. We use a zero-mean Gaussian prior that is scaled with an inverse variance hyperparameter $\alpha$, which results in 
\begin{equation}
     P(\pmb{\beta}\mid \alpha) = \prod_{i=1}^M \sqrt{\frac{\alpha}{2\pi}}\exp\left[-\frac{\alpha}{2}\beta_i^2\right].
    \label{eqn:tot_gaussian_prior}
\end{equation}
The MAP estimate and the covariance matrix can then be computed in closed form as
\begin{align}
    \pmb{\beta}_\mu &= (\mathbf{X}^\top\mathbf{X} + \alpha\sigma^2\mathbf{I})^{-1} \mathbf{X}^\top\mathbf{y},\\
    \pmb{\Sigma} &= \sigma^2 (\mathbf{X}^\top\mathbf{X} + \alpha\sigma^2\mathbf{I})^{-1}.   
    \label{eqn:covariance_matrix_sbr}
\end{align}
To induce sparsity in the derived model, an individual hyperparameter is introduced for each coefficient, which is re-estimated using the posterior mean and covariance by
\begin{align}
    \gamma_i &= 1 - \alpha_i\Sigma_{ii},\\
    \alpha_i^{new} &= \frac{\gamma_i}{\beta_{\mu,i}^2}.
\end{align}
Large values for $\alpha_i$ mean that there is high confidence that $\beta_i = 0$. Therefore, the corresponding basis functions $f_i$ can be dropped when $\alpha_i$ exceeds a certain threshold. \\
The probabilistic representation of the coefficients can now be utilized in various ways, particularly in the context of optimization \cite{Jones1998}. A widely used approach in Bayesian optimization is to define an acquisition function that indicates where the next sample should be placed so that as much information as possible about the system and its optimal solution is gathered. Popular examples of such acquisition functions are the expected improvement \cite{Mockus1978} and the probability of improvement \cite{Kushner1964}. These functions exhibit high values when either the predicted value is low or when the variance is high. Therefore, they balance exploration, which refers to searching in the region of the currently best solution, and exploitation, which refers to searching in areas that are coarsely sampled and show a high uncertainty. In the framework proposed in this work, using acquisition functions for the optimization is not appropriate, however, because the Bayesian regression model is formulated in terms of the entries of the reduced system matrices and not the objective function. We therefore resort to a different approach in Bayesian optimization, namely Thompson sampling (TS) \cite{Thompson1933}. The idea of this approach is to draw one realization from the posterior distribution of the Gaussian process and treat this as the true prediction of the reduced system matrices. This way, the proposed approach aligns with standard matrix interpolation frameworks, but also accounts for the uncertainty in the prediction of the reduced system matrices, especially when the pROM is built from a small number of samples.

\section{Gradient-based Optimization}
\label{sec:GB-Optimization}

A general optimization problem can be stated as:
\begin{equation}
\mathcal{P} \quad
\begin{cases}
    \begin{aligned}
        \min_{\mathbf{p}} \quad & R(\mathbf{p}), \\
        \text{s.t.} \quad & q_i(\mathbf{p}) \leq 0, \quad i = 1, \dots, n_q \\
        & h_j(\mathbf{p}) = 0, \quad j = 1, \dots, n_h \\
        & \mathbf{p} \in \mathbb{R}^d.
        \label{eqn:primal_optimization_problem1}
    \end{aligned}
\end{cases}
\end{equation}
Here, $R(\mathbf{p})$ is the objective function or response that shall be minimized by altering the design variable vector $\mathbf{p}$ while satisfying all $n_q$ inequality and $n_h$ equality constraints $q_i(\mathbf{p})$ and $h_j(\mathbf{p})$, respectively. This results in a size optimization problem in the frequency domain, as presented in \cite{Venini2016}. In this work, the objective function shown in \Cref{eq:objective_k} is used, and no equality or inequality constraints are imposed. The range of the parameters is constrained by the bounded domain $\Omega \in \mathbb{R}^d$, therefore, the optimization problem reads
\begin{equation}
\mathcal{P} \quad
\begin{cases}
    \begin{aligned}
        \min_{\mathbf{p}} \quad & R(\p) = \sqrt[k]{\sum_{i}^{n_s}  (|\mathbf{g}\mathbf{x}(\p,s_{i})|)^{k}}, \\
        \text{s.t.} \quad \mathbf{p} \in \Omega.
        \label{eqn:primal_optimization_problem2}
    \end{aligned}
\end{cases}
\end{equation}
A whole research field is devoted to developing methods that efficiently solve the optimization problem shown in \Cref{eqn:primal_optimization_problem2}. For an overview of these methods, the reader is referred to \cite{Martins2022}. In the following, we focus on gradient-based methods as they are more efficient when the response is smooth, and its gradients can be computed. The general idea of these approaches is to follow the gradient of the objective function towards the minimum, where this gradient vanishes. It is therefore important to obtain gradient information of the objective, which shall be elaborated on in the next section.

\subsection{Sensitivity Analysis}
\label{sec:SensitivityAnalysis}

The simplest way to obtain the gradients required for the optimization is to perform finite differences, so computing the response for a small perturbation and building the difference. However, this approach is computationally expensive, as for each parameter, a reduced system would have to be solved for a neighboring point. To circumvent this, the local derivatives can be computed via sensitivity analysis. For a simpler notation, we will, in the following, refer to $\tilde{\mathbf{K}}_r$ as the reduced dynamic stiffness matrix, so $\tilde{\mathbf{K}}_r = \K_r + s\C_r + s^2\M_r$. Furthermore, we omit the dependence on $\p$ for better readability. We consider an objective function $R_r\left(\mathbf{p}, \mathbf{x}_r(\mathbf{p})\right)$ based on the reduced system that depends on the reduced displacement field and thus in turn on the parameters as shown in \Cref{eq:objective_k}. The local derivative of this function is
\begin{equation}
    \frac{\mathrm{d}R_r}{\mathrm{d}\mathbf{p}} = \frac{\partial{R}_r}{\partial{\mathbf{p}}} + \left(\frac{\partial{R}_r}{\partial{\mathbf{x}_r}}\right)^\intercal\frac{\mathrm{d}{\mathbf{x}_r}}{\mathrm{d}{\mathbf{p}}}.
    \label{eq:SA_Gradient}
\end{equation}
Deriving the reduced equation of motion $\tilde{\mathbf{K}}_r \mathbf{x}_r = \mathbf{f}_r$, leads to an expression for the derivative~$\frac{\mathrm{d}\mathbf{x}_r}{\mathrm{d}\mathbf{p}}$:
\begin{align}
    \tilde{\mathbf{K}}_r \mathbf{x}_r &= \mathbf{f}_r 
    \quad \quad \Bigg| \frac{\mathrm{d}}{\mathrm{d}\mathbf{p}} \label{eq:EoM_dynamicStiffness} \\
    \frac{\mathrm{d}}{\mathrm{d}\mathbf{p}}  \left( \tilde{\mathbf{K}}_r \mathbf{x}_r \right) &= \frac{\mathrm{d}\mathbf{f}_r}{\mathrm{d}\mathbf{p}}  \\
    \frac{\mathrm{d}\tilde{\mathbf{K}}_r}{\mathrm{d}\mathbf{p}} \mathbf{x}_r + \tilde{\mathbf{K}}_r \frac{\mathrm{d}\mathbf{x}_r}{\mathrm{d}\mathbf{p}} &= \frac{\mathrm{d}\mathbf{f}_r}{\mathrm{d}\mathbf{p}} \\
        \frac{\mathrm{d}\mathbf{x}_r}{\mathrm{d}\mathbf{p}} &= \tilde{\mathbf{K}}_r^{-1} \underbrace{\left( \frac{\mathrm{d}\mathbf{f}_r}{\mathrm{d}\mathbf{p}} - \frac{\mathrm{d}\tilde{\mathbf{K}}_r}{\mathrm{d}\mathbf{p}} \mathbf{x}_r \right)}_{\text{Pseudo load vector } \mathbf{f}_r^*}.
    \label{eq:dxdp}
\end{align}
This results in the following expression for the total gradient of the objective function:
\begin{equation}
    \frac{\mathrm{d}R_r}{\mathrm{d}\mathbf{p}} = \frac{\partial{R}_r}{\partial{\mathbf{p}}} +\underbrace{\frac{\partial{R}_r}{\partial{\mathbf{x}_r}} \, \tilde{\mathbf{K}}_r^{-1}}_{\text{Adjoint variables } \boldsymbol{\eta}^\top} \left( \frac{\mathrm{d}\mathbf{f}_r}{\mathrm{d}\mathbf{p}} - \frac{\mathrm{d}\tilde{\mathbf{K}}_r}{\mathrm{d}\mathbf{p}} \mathbf{x}_r \right).
    \label{eq:SA_Gradient2}
\end{equation}
In case there are more parameters than objective functions, it is more efficient to first solve for the so-called adjoint variable via
\begin{equation}
    \tilde{\mathbf{K}}_r^\intercal \boldsymbol{\eta}  = \left( \frac{\partial{R}_r}{\partial{\mathbf{x}_r}} \right)^\intercal,
\end{equation}
and then multiply this by the pseudo-load vector for each parameter. Computing the local derivative of the reduced objective function then only requires solving a reduced system once per objective function. The derivatives of the reduced input vector and the reduced dynamic stiffness matrix can be directly computed from the analytical expressions that are available due to the matrix interpolation. According to the regression formula outlined in \Cref{eq:Regression}, the derivative of an entry $a$ of the reduced operators is given by
\begin{equation}
    \frac{\partial a}{\partial \p} = \sum_{i=1}^{n_f} \frac{\partial f_i}{\partial \p} \beta_i.
    \label{eq:deriv_polynomial}
\end{equation}
The derivatives $\frac{\partial{R}_r}{\partial{\mathbf{p}}}$ and $\frac{\partial{R}_r}{\partial{\mathbf{x}_r}}$ can be computed beforehand based on the chosen objective function. Thereby, the first is usually zero, while the latter is more intricate. This is because the objective function is a real-valued function but depends on the reduced vector of degrees of freedom $\x_r$, which is a complex variable for damped systems. Such functions are not holomorphic and thus not complex differentiable. In order to compute the derivative of non-holomorphic functions, the so-called \textit{Wirtinger operators} \cite{Wirtinger1927} are required. These treat a complex variable $z = x + \mathrm{i}y$ and its complex conjugate $\bar{z}$ in the derivative like separate variables, so that they write
\begin{align}
\frac{\partial}{\partial z} &:= \frac{1}{2} \left[ \frac{\partial}{\partial x} - \mathrm{i} \frac{\partial}{\partial y} \right], \text{ and } \label{eqn:wirtinger_derivs1} \\ 
\frac{\partial}{\partial \bar{z}} &:= \frac{1}{2} \left[ \frac{\partial}{\partial x} + \mathrm{i} \frac{\partial}{\partial y} \right].\label{eqn:wirtinger_derivs2}
\end{align}
The final differential $\mathrm{d}f$ of a real-valued function $f$ with complex arguments $f(z) : \mathbb{C} \rightarrow \mathbb{R}$ then reads as
\begin{equation}
    \mathrm{d}f = 2 \cdot \Re \left[ \frac{\partial{f(z)}}{\partial{z}} \mathrm{d} z \right],
    \label{eq:wirt_2Re}
\end{equation}
where $\Re(\cdot)$ refers to the real part. The derivative of the absolute value of a complex number, for example, thus results in
\begin{equation}
    \frac{\mathrm{d}|z|}{\mathrm{d}z} = \Re \left[ \frac{\bar{z}}{|z|} \right]. \label{eq:Wirtinger}
\end{equation}
Before providing the final derivative of the objective function given in \Cref{eq:objective_k}, one further step is necessary. In \Cref{eq:objective_k}, the response is computed as the $L_k$-norm of $n_s$ discrete outputs referring to discrete frequencies $s_i$, $i = 1, \dots, n_s$ in a specific frequency range. For computing the sensitivities of the system, \Cref{eq:EoM_dynamicStiffness} is rewritten in terms of $n_s$ subproblems of the type
\begin{equation}
\tilde{\mathbf{K}}(s_i) \cdot \mathbf{x}(s_i) = \mathbf{f}(s_i), \quad i = 1, \dots, n_s.
\end{equation}
Concatenating all subproblems, the following system of equations is obtained:
\begin{equation}
\tilde{\mathbf{K}}_{\text{tot}} \cdot \mathbf{x}_{\text{tot}} = \mathbf{f}_{\text{tot}}.
\label{eqn:tot_sys}
\end{equation}
There,
\begin{equation}
\tilde{\mathbf{K}}_{\text{\text{tot}}} = \begin{bmatrix} \tilde{\mathbf{K}}(s_1) & 0 & \cdots & 0 \\ 0 & \tilde{\mathbf{K}}(s_2) & \cdots & 0 \\ \vdots & \vdots & \ddots & \vdots \\ 0 & 0 & \cdots & \tilde{\mathbf{K}}(s_{n}) \end{bmatrix}, \quad \mathbf{x}_{\text{tot}} = \begin{bmatrix} \mathbf{x}(s_1) \\ \mathbf{x}(s_2) \\ \vdots \\ \mathbf{x}(s_{n}) \end{bmatrix}, \text{ and } \quad \mathbf{f}_{\text{tot}} = \begin{bmatrix} \mathbf{f}(s_1) \\ \mathbf{f}(s_2) \\ \vdots \\ \mathbf{f}(s_{n}) \end{bmatrix}. 
\end{equation}
With this, the derivative of the objective function given in \Cref{eq:objective_k} can be computed as
\begin{equation}
    \frac{\mathrm{d}R}{\mathrm{d}\mathbf{p}} = \Re \Bigg(\sum_{i}^{n_s} \left[\left( \sum_{j=1}^{n_s} |y_j|^k \right)^{\frac{1}{k} - 1} \cdot |y_i|^{k-1} \cdot \frac{\bar{y_{i}}}{  |y_{i}|} \cdot \begin{bmatrix} \mathbf{0}  \cdots  \mathbf{g}_i  \cdots  \mathbf{0} \end{bmatrix}^\top\right] \cdot \tilde{\mathbf{K}}_{\text{tot}}^{-1}\left( - \frac{\mathrm{d}\tilde{\mathbf{K}}_{\text{tot}}}{\mathrm{d}\mathbf{p}} \mathbf{x}_{\text{tot}}\right)\Bigg).
\end{equation}
A detailed derivation of this derivative is shown in \Cref{sec:appendix_gradient}.

\section{Methodology}%
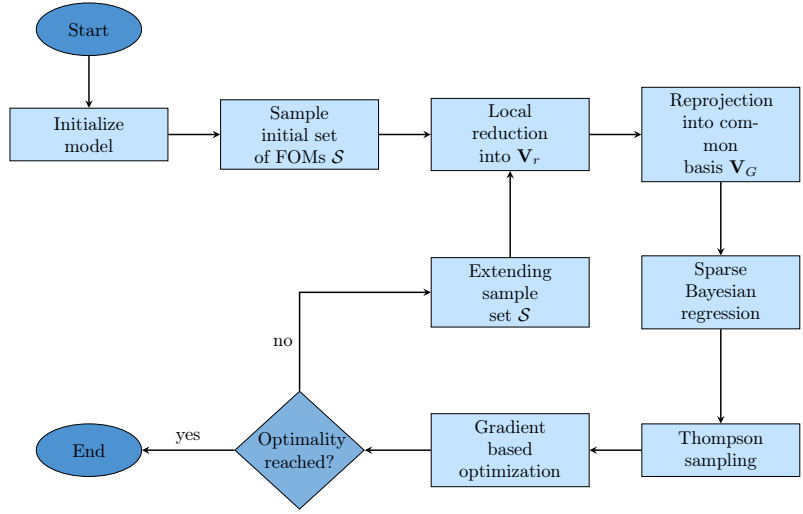
\begin{figure}[H]
    \centering
    \scalebox{0.7}{
        \begin{tikzpicture}[node distance=2cm]
            \node (start) [startstop] {Start};
            \node (init) [process, below of=start] {Initialize model};
            \node (sample) [process, right of=init, node distance = 4 cm] {Sample initial set of FOMs $\mathcal{S}$};
            \node (reduction) [process, right of=sample, node distance = 4 cm] {Local reduction into $\mathbf{V}_r$};
            \node (reprojection) [process, right of=reduction, node distance = 4 cm] {Reprojection into common basis $\mathbf{V}_G$};
            \node (training) [process, below of=reprojection, node distance = 3 cm] {Sparse Bayesian regression};
            \node (ts) [process, below of=training, node distance = 3 cm] {Thompson sampling};
            \node (optimization) [process, left of=ts, node distance = 4 cm] {Gradient based optimization};
            \node (decision) [decision, left of=optimization, node distance = 4 cm] {Optimality reached?};
            \node (add) [process, left of=training, node distance = 4 cm] {Extending sample set $\mathcal{S}$};
            \node (end) [startstop, left of=decision, node distance = 4cm] {End};
            \draw [arrow] (start) -- (init);
            \draw [arrow] (init) -- (sample);
            \draw [arrow] (sample) -- (reduction);
            \draw [arrow] (reduction) -- (reprojection);
            \draw [arrow] (reprojection) -- (training);
            \draw [arrow] (training) -- (ts);
            \draw [arrow] (ts) -- (optimization);
            \draw [arrow] (optimization) -- (decision);
            \draw [arrow] (decision.north) |- node[pos=0.25, left]{no} (add.west);
            \draw [arrow] (add) -- (reduction);
            \draw [arrow] (decision) -- node[midway, above]{yes} (end);
            
        \end{tikzpicture}
        }
    \caption{Flowchart of the adaptive sampling algorithm.}
    \label{fig:Algorithm Flowchart}
\end{figure}

\begin{figure}[H]
	\centering
		 {\includegraphics[ width=1\textwidth]{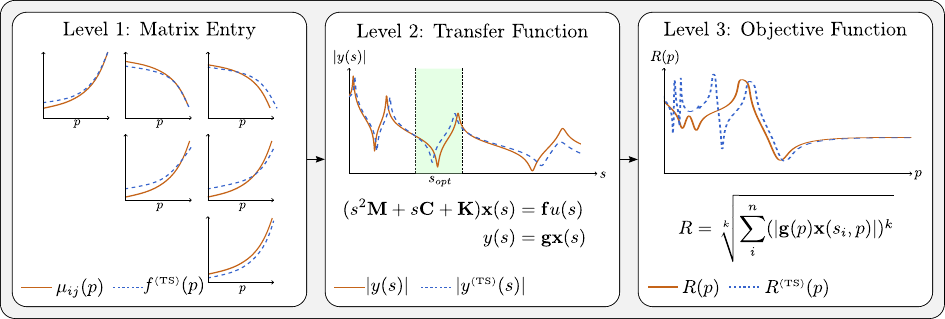}}
	\caption{Overview of how uncertainties are propagated throughout the workflow.}
 \label{fig:Thompson_Sampling_3Levels}
\end{figure}

\label{sec:methodology}
In Figure \ref{fig:Algorithm Flowchart}, the outline of the adaptive sampling workflow is provided. The fundamental goal is to deploy a gradient-based optimizer that works on the ROM level and uses
analytic adjoint sensitivity analysis to determine an optimum of the objective function defined in \Cref{eq:objective_k}. The identified parameter configuration is appended to the sample set and will serve as the next observation, i.e., the next sampled FOM. By deploying the fitness itself as the acquisition criterion, the parametric model is refined locally in promising regions while sampling effort in less promising regions is reduced. 
\textbf{Initial sampling and reduction} $\,$ Starting with a rather sparse initial sampling, say a $3^d$ full factorial design, each of the collected FOMs is reduced locally using IRKA as described in Section \ref{sec:pMOR}. In order to allow a meaningful interpolation between the reduced operators of the system, the reduction bases need to be adjusted accordingly. As described in Section \ref{sec:pMOR_MI}, this issue is tackled as proposed in \cite{Degroote2010} by determining one common basis by a singular value decomposition of all concatenated bases, and reprojecting the FOMs into this common basis (see Algorithm \ref{alg:global_basis}). By that, we omit the need for additional, subregion-specific basis information in order to describe and evaluate the parametrized ROM. 
\begin{algorithm}[H]
\caption{Global basis computation}
\begin{algorithmic}[1]
\Require Locally reduced bases $\mathbf{V}_{r,i}$, SVD retention fraction $\kappa$
\Ensure Global basis $\mathbf{V}_G$, Updated ROM order $r$
\Function{GlobalBasisComputation}{$\mathbf{V}_r$, $\kappa$}
    \State Concatenate all local bases: \quad $\mathbf{V}_{\text{all}} \gets \text{concatenate}(\mathbf{V}_{r,i})$
    \State Compute SVD: \quad $[\mathbf{U}, \boldsymbol{\Sigma}, \sim] \gets \text{SVD}(\mathbf{V}_{\text{all}})$
    \State Determine truncation index $r$:
    \State \quad $\boldsymbol{\sigma} \gets \text{diag}(\boldsymbol{\Sigma})$ \quad $c \gets \text{cumsum}(\boldsymbol{\sigma})$\quad  $r \gets \min\{j : c_j/c_{end} > \kappa\}$
    \State Extract global basis: \quad $\mathbf{V}_G \gets \mathbf{U}_{:,1:r}$
    \State Reproject system matrices in global basis: \quad $\mathbf{K}_{i,r} \gets \mathbf{V}_G^\top \mathbf{K}_i \mathbf{V}_G$\\
\Return $\mathbf{V}_G$, $\mathbf{K}_r$, $\mathbf{D}_r$, $\mathbf{M}_r$, $\mathbf{f}_r$, $\mathbf{g}_r, r$
\EndFunction
\end{algorithmic}
\label{alg:global_basis}
\end{algorithm}
\textbf{Operator interpolation} $\,$ After projecting all FOMs into the same reduced basis, an interpolation of the reduced operators can be performed in order to evaluate the model for arbitrary parameter configurations. This surrogate model allows the subsequent optimization process to be conducted
with significantly reduced computational cost compared to the FOM. In the simplest case, this surrogate model can be a multivariate polynomial regression, which would yield a separate polynomial function for each matrix entry. That however, would result in a highly exploitative algorithm which would be prone to converging into local optima. In order to introduce some exploration into the sample acquisition strategy, the simple polynomial regression is replaced by a sparse Bayesian regression (SBR) routine provided by \cite{tipping2001}, with a subsequent Thompson sampling of the polynomial coefficients. Instead of yielding deterministic values for the polynomial coefficients of the regression models, SBR yields a probabilistic distribution for each entry consisting of the mean value $\boldsymbol{\beta}_\mu$ and the covariance matrix $\boldsymbol{\Sigma}$ (see Section \ref{sec:sbr}). By Thompson-sampling from these distributions, the uncertainty of the learned coefficients describing the matrix entries is accounted for and will be propagated through the three main levels of the algorithm as illustrated in \Cref{fig:Thompson_Sampling_3Levels}. Especially in the early stages of the algorithm, when samples are only sparsely distributed, this extension will make the objective function resulting from the Thompson-sampled coefficients deviate from the objective obtained from the mean coefficients. Therefore, the optimizer will converge to regions of high uncertainty more often. 

\textbf{Gradient calculation and optimization} $\,$ MATLAB's \lstinline{fmincon} optimizer is used to perform a sizing optimization that minimizes the defined objective function. The learned models are leveraged to determine the analytical gradients of the objective function as described in Section \ref{sec:GB-Optimization}. In order to do so, the partial derivative of the objective function with respect to the state vector $\frac{\partial{R}}{\partial{\mathbf{x}}}$ has to be evaluated as laid out in \Cref{eq:SA_Gradient2} and Appendix \ref{sec:appendix_gradient}. Furthermore, the obtained polynomial models with the Thompson-sampled coefficients $f^{\text{(TS)}}(p)$ can be differentiated using simple calculus shown in \Cref{eq:deriv_polynomial} in order to obtain the derivative of the reduced parametrized operators with respect to the parameters $\frac{\mathrm{d}\tilde{\mathbf{K}}}{\mathrm{d}\mathbf{p}}$. The pseudo code for evaluating the gradient at a given parameter point is given in \Cref{alg:grad}. \\

\textbf{Sample set extension} $\,$ The identified optimum is added to the sample set and will be used to refine the parametrized ROM iteratively. In order to avoid very close clustering, a simple distance check is performed, which moves the new sample away from its closest neighbor if a certain threshold is undercut. As a stopping criterion, the relative improvement of the objective within the last two
iterations is calculated. Once this improvement reaches below a certain threshold or if
the maximum number of iterations is reached, the optimization is terminated.

\begin{algorithm} [H]
\caption{Gradient evaluation}
    \begin{algorithmic}[1]
    \Require 
        \State Total system operators $\tilde{\mathbf{K}}_{\text{tot}}$, $\mathbf{x}_{\text{tot}}$, $\mathbf{g}_{\text{tot}}$
        \State Parameter point $\hat{\mathbf{p}}$, frequency points $f_{\text{opt}}$
        \State Trained Thompson sampled models $\text{Mdl\_K}$, $\text{Mdl\_D}$, $\text{Mdl\_M}$, cost value $R$
    \Ensure Gradient $\mathrm{d}R/\mathrm{d}\mathbf{p}$
    
\Function{EvaluateGradient}{$\tilde{\mathbf{K}}_{\text{tot}}$, $\mathbf{x}_{\text{tot}}$, $\mathbf{g}_{\text{tot}}$, $\hat{\mathbf{p}}$, $f_{\text{opt}}$, $k$}
        \State Build derivatives of regression models: \quad $\text{Mdl\_dKdp} \gets \text{GetMatrixDerivative}(\text{Mdl\_K})$
        \State Evaluate matrix derivatives: \quad  $\frac{\mathrm{d}\mathbf{K}}{\mathrm{d}\mathbf{p}} \gets \text{EvaluateMatrixDerivative}(\text{Mdl\_dKdp}, \hat{\mathbf{p}})$
        \State Calculate complex frequency: \quad $s \gets 2\pi\mathrm{i}\,f_{\text{opt}}$
        \State Assemble dynamic stiffness matrix derivative 
        \For{\text{all $s$ in $s_{\text{opt}}$}}
            \State $\frac{\mathrm{d}\tilde{\mathbf{K}}}{\mathrm{d}\mathbf{p}} \gets \frac{\mathrm{d}\mathbf{K}}{\mathrm{d}\mathbf{p}} + s\frac{\mathrm{d}\mathbf{D}}{\mathrm{d}\mathbf{p}} + s^2\frac{\mathrm{d}\mathbf{M}}{\mathrm{d}\mathbf{p}}$
        \EndFor
        \State Compute derivative of response w.r.t. state vector $\mathrm{d}R/\mathrm{d}\mathbf{x}$ (see Equation (\ref{eqn:dRdx_damped}))
        \State Assemble gradient:
        \State $\pmb{\eta}^\top \gets (\mathrm{d}R/\mathrm{d}\mathbf{x})\tilde{\mathbf{K}}_{\text{tot}}^{-1}$
        \State $\mathrm{d}R/\mathrm{d}\mathbf{p} \gets \pmb{\eta}^\top(-\frac{\mathrm{d}\tilde{\mathbf{K}}}
        {\mathrm{d}\mathbf{p}}\mathbf{x}_{\text{tot}})$
        \State $\mathrm{d}R/\mathrm{d}\mathbf{p} \gets \frac{1}{R} \cdot \Re(\mathrm{d}R/\mathrm{d}\mathbf{p})$\\
    \Return $\mathrm{d}R/\mathrm{d}\mathbf{p}$
\EndFunction
\end{algorithmic}
\label{alg:grad}
\end{algorithm}

\section{Results}
\label{sec:Results}

In the following section, the numerical examples and the results are summarized and presented using two example systems: A cantilever Timoshenko beam and the Kelvin cell structural topology. The implementation was carried out using the beam example due to the low computational effort of the corresponding FOM as well as the relatively simple dynamics of the system. Subsequently, the algorithm was enhanced and validated on the Kelvin cell, a computationally more demanding model with a more complex dynamic.

\subsection{Timoshenko Beam}
\begin{figure} [H]
    \begin{minipage}[t]{0.35\textwidth}
    \vspace{0pt}  
        \centering
        \includegraphics[width=0.9\linewidth]{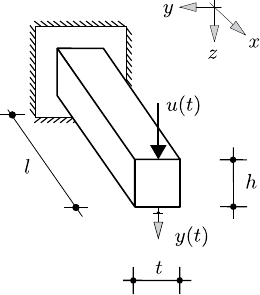}
    \end{minipage}%
    \begin{minipage}[t]{0.5\textwidth}
    \centering
        \vspace{0.3em}
        \resizebox{\linewidth}{!}{%
        \begin{tabular}{llll}
            \hline
            Parameter & Description & Value & Unit \\
            \hline
            \multicolumn{4}{l}{\textbf{Geometric parameters}} \\
            $l$ & Beam length & $1.0$ & m \\
            $t$ & Beam thickness & $[0.01,0.05]$ & m \\
            $h$ & Beam height & $[0.01,0.05]$ & m \\
            $N$ & Number of elements & 400 & - \\
            DOFs & Degrees of freedom & $12N$ & - \\
            \hline
            \multicolumn{4}{l}{\textbf{Material properties}} \\
            $\rho$ & Density & $7.86 \cdot 10^3$ & kg/m$^3$ \\
            $E$ & Young's modulus & $2.10 \cdot 10^{11}$ & N/m$^2$ \\
            $\nu$ & Poisson's ratio & 0.3 & - \\
            $G$ & Shear modulus & $\frac{E}{2(1+\nu)}$ & N/m$^2$ \\
            \hline
            \multicolumn{4}{l}{\textbf{Damping parameters}} \\
            $\alpha$ & Rayleigh damping & $8$ & s \\
            $\beta$ & Rayleigh damping & $8 \cdot 10^{-6}$ & 1/s \\
            \hline
            \end{tabular}
            }
\end{minipage}
    \caption{Timoshenko beam model and relevant parameters}
    \label{fig:Beam_with_table}
\end{figure}

\begin{figure}[H]
    \centering
    \begin{subfigure}[b]{\textwidth}
        \centering
            \includegraphics[width=0.6\textwidth,
  trim = 0cm 11.3cm 0cm 0cm,
  clip]{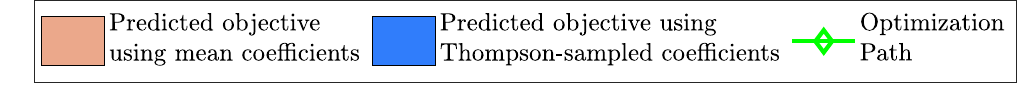}
    \end{subfigure}
    \vspace{0.4cm} 
    \begin{subfigure}[b]{0.32\textwidth}
        \centering
        \adjustbox{trim=0cm 0cm 0cm 0cm,clip}{
            \includegraphics[width=\textwidth]{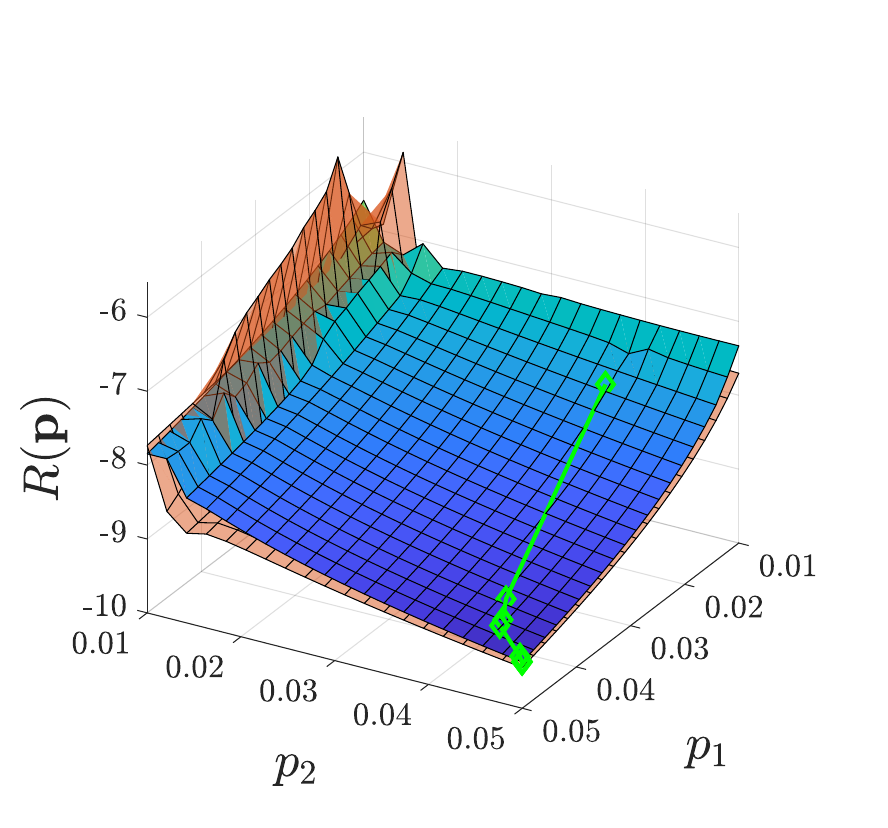}
        }
        \caption{Iteration 1}
        \label{fig:Beam2D_Iteration01}
    \end{subfigure}
    \hspace{-0cm}  
    \begin{subfigure}[b]{0.32\textwidth}
        \centering
        \adjustbox{trim=0cm 0cm 0cm 0cm,clip}{
            \includegraphics[width=\textwidth]{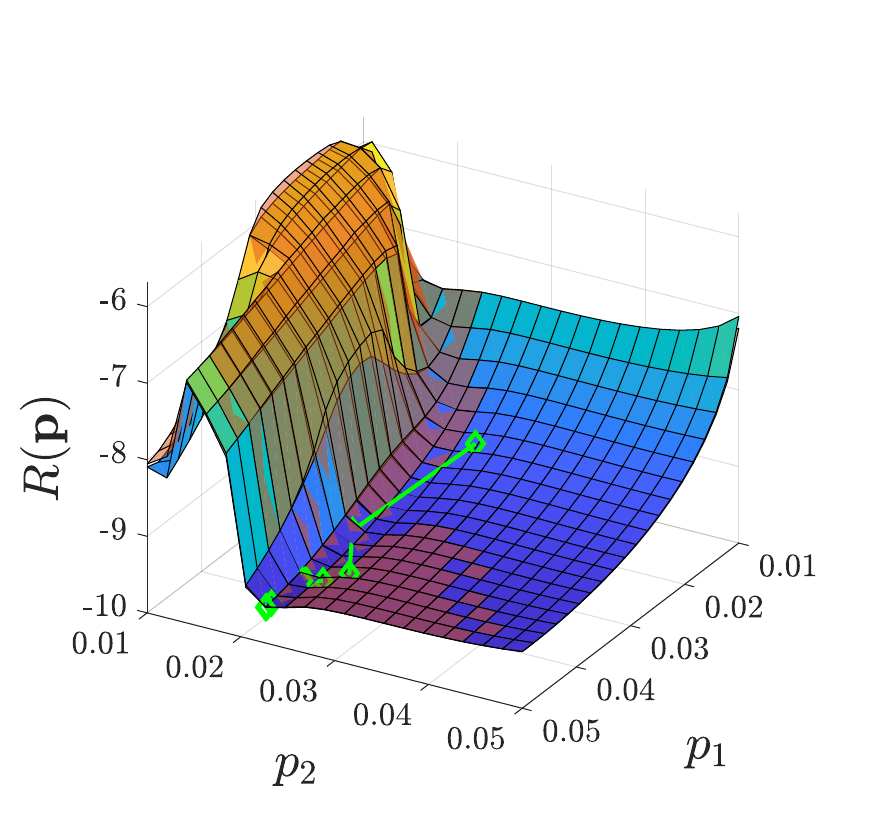}
        }
        \caption{Iteration 5}
        \label{fig:Beam2D_Iteration05}
    \end{subfigure}
    \hspace{-0cm}  
    \begin{subfigure}[b]{0.32\textwidth}
        \centering
        \adjustbox{trim=0cm 0cm 0cm 0cm,clip}{
            \includegraphics[width=\textwidth]{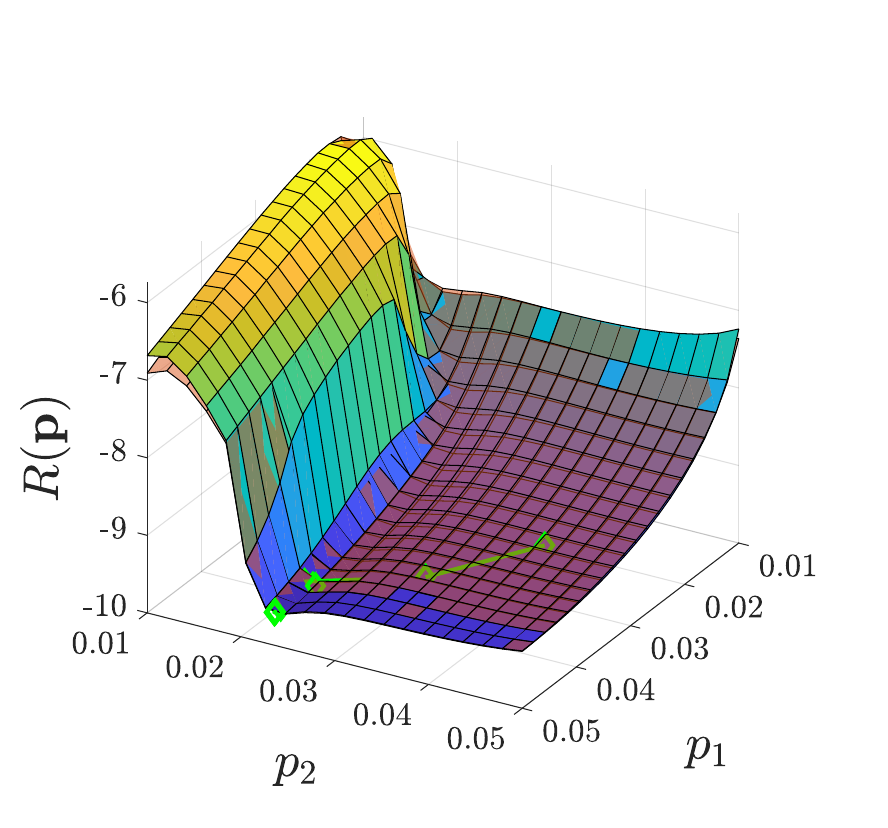}
        }
        \caption{Iteration 7}
        \label{fig:Beam2D_Iteration07}
    \end{subfigure}
    \caption[Selected iterations of the objective function refinement for the 2D Timoshenko beam example]{Selected iterations of the objective function refinement for the 2D Timoshenko beam example.}
    \label{fig:Beam_results}
\end{figure}

The Timoshenko beam model, as depicted in Figure \ref{fig:Beam_with_table}, is generated using the MATLAB code provided by \cite{Panzer2009} using 400 Timoshenko beam elements with twelve local degrees of freedom each. A downward force is applied to the tip of the beam. For the output, the displacement of the tip of the beam is examined. The used material and geometric parameters are listed below. In the two-dimensional case, both height $h$ and thickness $t$ are varied. The sampled FOMs are reduced into local bases $\mathbf{V}_r$ of size $r = 6$, which are then used to calculate the global basis $\mathbf{V}_G$, as specified in Algorithm \ref{alg:global_basis}. After this global basis was found, all FOMs are reduced as described previously. Using the reduction parameters specified in Table \ref{tab:beam_rom_parameters}, the final global basis contains $r = 10$ basis vectors for this example.
\begin{table}[H]
    \centering
    \caption{MOR parameters for the Timoshenko beam example}
    \begin{tabular}{llll}
        \hline
        Parameter & Description & Value & Unit \\
        \hline
        $r$ & Reduced order model size & 6 & - \\
        $\varepsilon_{\text{IRKA}}$ & IRKA tolerance & $1 \times 10^{-6}$ & rad/s \\
        $N$ & Maximum IRKA iterations & 10 & - \\
        $f_{0, \text{IRKA}}$ & Initial expansion frequencies & linspace($-250$, $250$, $r$) & 1/s \\
        $\kappa_{SVD}$ & Retained information after re-projection & 0.9995 & - \\
        \hline
    \end{tabular}
    \label{tab:beam_rom_parameters}
\end{table}
For the two-dimensional Timoshenko beam example, the starting sample distribution is chosen as coarsely as possible in order to allow a simple linear interpolation of data, i.e., a $2^2$ full-factorial design (FFD). The model's transfer function ought to be optimized in the frequency range of $f_\text{opt} = [50, 100]$ Hz with the order of the norm $k=1$. The results of a representative run are shown in Figure \ref{fig:Beam_results}. The algorithm first repeatedly finds a local optimum at $\boldsymbol{p} = [p_1, p_2] =  [t,h] \approx [0.05, 0.05]$ due to the very poor approximation provided by a linear interpolation of the reduced operators. The insufficient order of the polynomials fitted through the reduced operator entries essentially makes the model unable to properly capture the relationship between the design variables and the transfer function behavior, resulting in a poorly approximated objective function (see Figure \ref{fig:Beam2D_Iteration01}).  Even though this sets a total of three new samples close to the already existing sample at the boundary, these samples still
help improve the regression models so that after the third iteration, a shape similar to the true shape of the objective develops for the first time. In the following iterations, the objective function is further refined, and a total of four samples are set around
the area of the optimum at $\boldsymbol{p} \approx [0.05, 0.0244]$. After seven additional samples are set, the algorithm terminates with a total number of 11 samples. The final sample distribution and the iteration details can be found in Figure \ref{fig:Beam_results}.
\begin{figure}[H]
    \centering
    \begin{minipage}[t]{0.5\textwidth}
        \vspace{-3.3cm}
        \hspace{0cm}
        \centering
        \adjustbox{trim=0cm 0cm 0cm 0cm,clip}{
        \includegraphics[width=\textwidth]{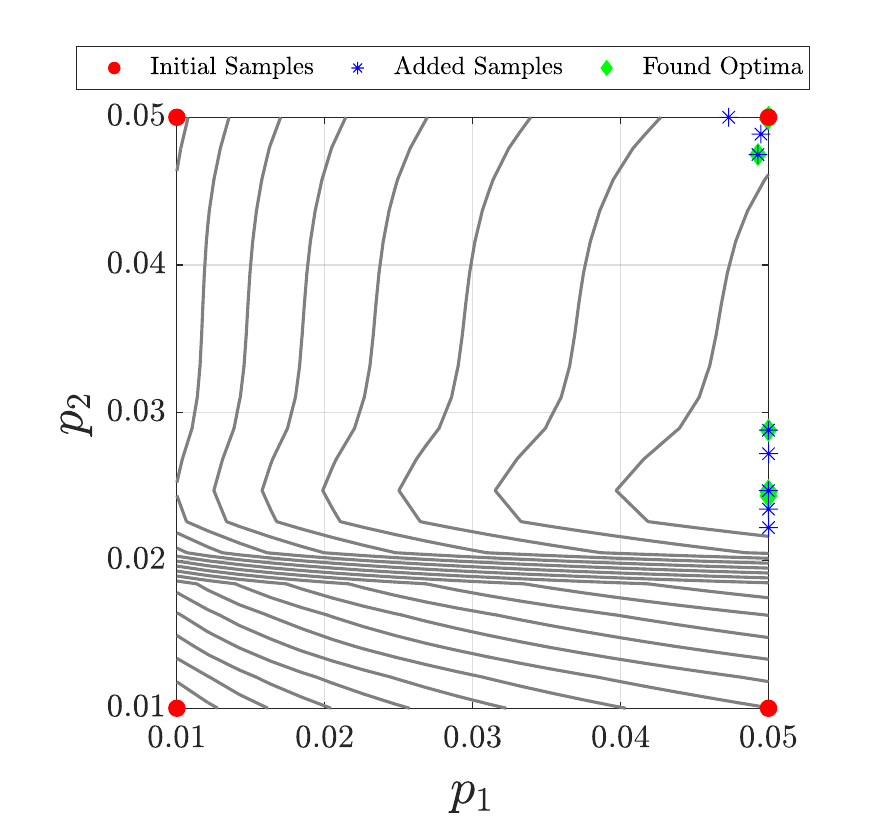}}
    \end{minipage}%
    \begin{minipage}[t]{0.5\textwidth}
        \hspace{-1.5cm}
        \vspace{0cm}
        \centering
        \adjustbox{scale=0.8}{%
        \begin{tabular}{lc|c|c}
            \Xhline{1.5pt}
            \multicolumn{2}{l|}{\makecell{Initial samples}} & \multicolumn{2}{c}{$4$ samples , $2^2$ FFD}\\
            \hline
            \multicolumn{2}{l|}{\makecell{True optimum}} & \multicolumn{2}{c}{$p^* = [t^*, h^*] = [0.05, 0.02437]$ m} \\
            \hline
            \multicolumn{2}{l|}{\makecell{Objective}} & \multicolumn{2}{c}{$f_{\text{opt}} \in [50,100] \, \text{Hz}$, $k = 1$} \\
            \hline
            \multicolumn{2}{l|}{\makecell{Minimum \\distance treshold}} & \multicolumn{2}{c}{$12.5\times10^{-4}$ m} \\
            \Xhline{1.5pt}
            \multicolumn{2}{c|}{\makecell{Iteration}} & \makecell{Found\\optimum $\mathbf{p}^*_i$} & \makecell{Objective\\value $R(\mathbf{p}_i^*)$} \\
            \Xhline{1.5pt}
            \multirow{7}{*}\makecell{} 
            & $1$ & $[0.05, 0.05]$    & $-9.6110$  \\       \cline{2-4}
            & $2$ & $[0.05, 0.05]$    & $-9.3910$ \\       \cline{2-4}
            & $3$ & $[0.05, 0.05]$    & $-9.3715$  \\       \cline{2-4}
            & $4$ & $[0.05, 0.0223]$  & $-9.4784$  \\        \cline{2-4}
            & $5$ & $[0.05, 0.0234]$  & $-9.5270$  \\       \cline{2-4}
            & $6$ & $[0.05, 0.0241]$  & $-9.5354$  \\       \cline{2-4}
            & $7$ & $[0.05, 0.0244]$  & $-9.5991$  \\       \cline{2-4}
            & $8$ & $[0.05, 0.0244]$  & $-9.5990$  \\
            \Xhline{1.5pt}
            \multicolumn{2}{l|}{\makecell{Final samples}} & \multicolumn{2}{c}{$11$} \\
            \Xhline{1.5pt}
            \end{tabular}
            }
    \end{minipage}
    \caption[Optimization summary and final sample distribution for the 2D Timoshenko beam example]{Optimization summary and final sample distribution for the 2D Timoshenko beam example.}
    \label{fig:Beam2D_Final}
\end{figure}
\subsection{Kelvin cell}
\label{subsec:KC_results}

\begin{figure}[H]
    \centering
    \hspace{-0.5cm}
    \begin{minipage}{0.4\linewidth}
        \centering
        \vspace{1cm}
		\includegraphics[width=\textwidth]{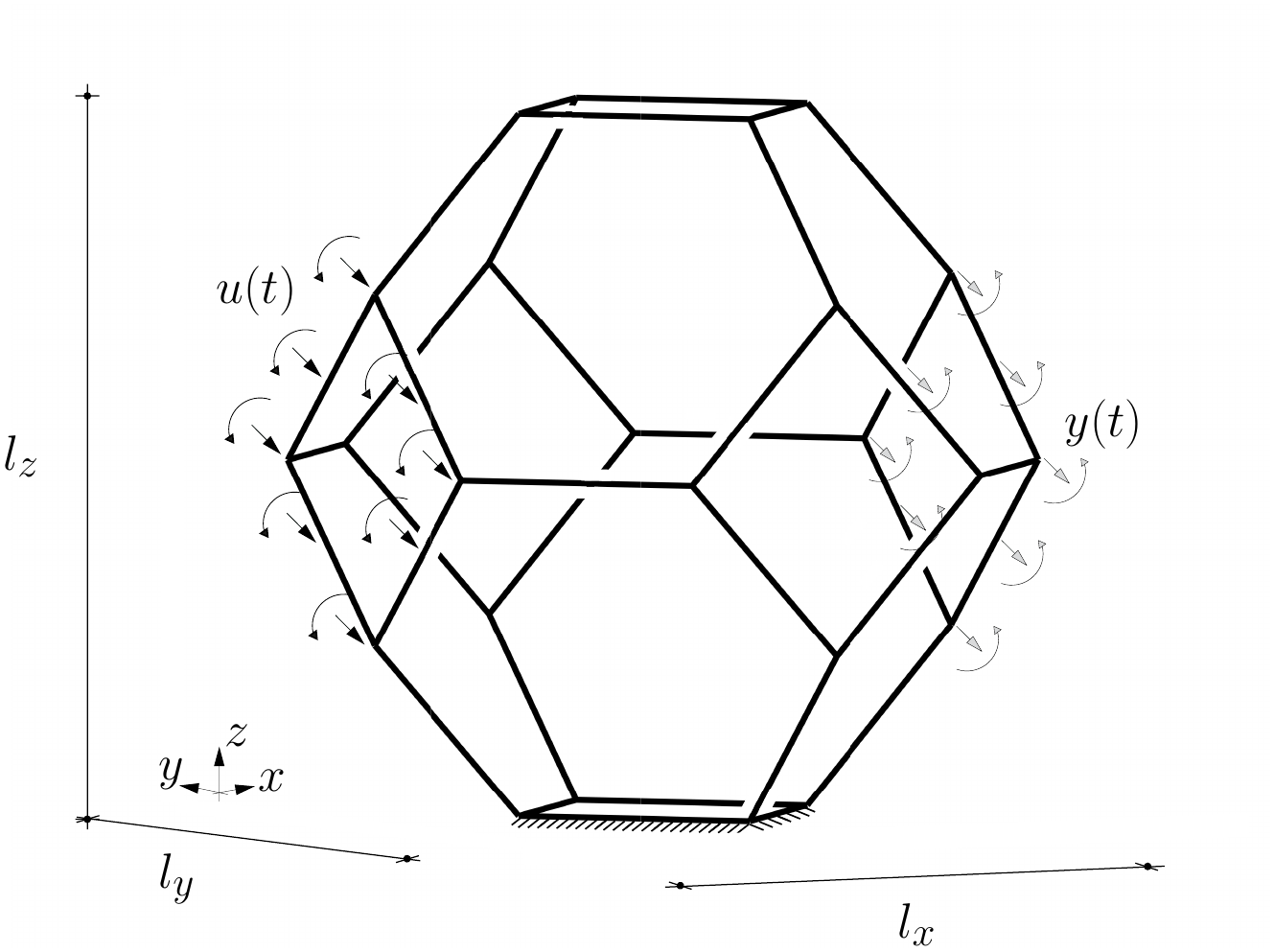}
        \label{fig:KelvinCell}
    \end{minipage}
    \hspace{0cm}
    \begin{minipage}{0.5\linewidth}
        \centering
        \vspace{0cm}
        \begin{adjustbox}{width=\linewidth}
            \begin{tabular}{llll}
                \hline
                Parameter & Description & Value & Unit \\
                \hline
                \multicolumn{4}{l}{\textbf{Geometric parameters}} \\
                $l_x$ & Length & $[0.055,0.080]$ & m \\
                $l_y$ & Length & $[0.020,0.045]$ & m \\
                $l_z$ & Length & $0.05$ & m \\
                $t$ & Beam thickness & $0.001$ & m \\
                $N$ & Number of elements & 1800 & - \\
                DOFs & Degrees of freedom & $10800$ & - \\
                \hline
                \multicolumn{4}{l}{\textbf{Material properties}} \\
                $\rho$ & Density & $1.18 \cdot 10^3$ & kg/m$^3$ \\
                $E$ & Young's modulus & $4.35 \cdot 10^{9}$ & N/m$^2$ \\
                $\nu$ & Poisson's ratio & 0.3 & - \\
                $G$ & Shear modulus & $\frac{E}{2(1+\nu)}$ & N/m$^2$ \\
                \hline
                \multicolumn{4}{l}{\textbf{Damping parameters}} \\
                $\alpha$ & Rayleigh damping & $8 \cdot 10^{-6}$ & 1/s \\
                $\beta$ & Rayleigh damping & $8$ & s \\
                \hline
            \end{tabular}
        \end{adjustbox}
        
    \end{minipage}
    \caption{Kelvin cell model and relevant parameters
    \label{fig:KelvinCell_parameters}
}
\end{figure}

The Kelvin cell model is depicted in Figure \ref{fig:KelvinCell_parameters}. The cell is clamped at the bottom struts while an input force is applied to the left side. For the output, the mean displacement of the right side of the cell is taken. 50 square Timoshenko beam elements are used to discretize each strut of the cell, resulting in approximately 1,800 elements and 10,800 DOFs for the FOM. The table shows the relevant parameters for the generation of the FE model. For the optimization, the two lengths $l_x$ and $l_y$ are varied. 

\begin{figure}[H]
    \centering
    \hspace{0cm}
    \begin{minipage}{0.5\linewidth}
        \centering
        \vspace{0cm}
        \includegraphics[width=1\linewidth, trim=0cm 0cm 0cm 0cm,clip]{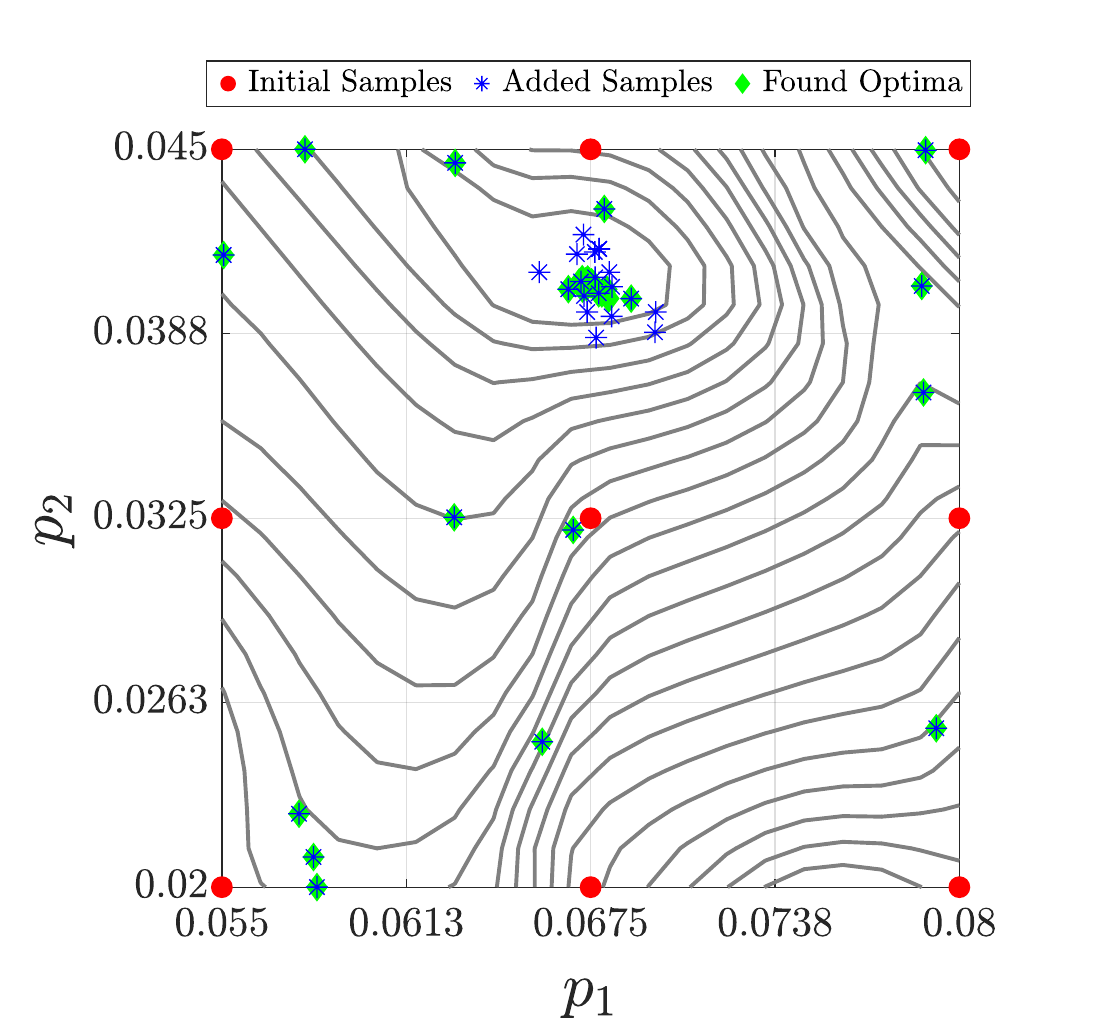}
        \label{fig:KC_2D_3x3_Final_Distribution_CheckDistance}
    \end{minipage}
    \hspace{-0.9cm}
    \begin{minipage}{0.5\linewidth}
        \centering
        \vspace{-0.4cm}
        \begin{adjustbox}{width=1.18\linewidth}
            \includegraphics[width=\linewidth, trim=0cm 0cm 0cm 0cm,clip]{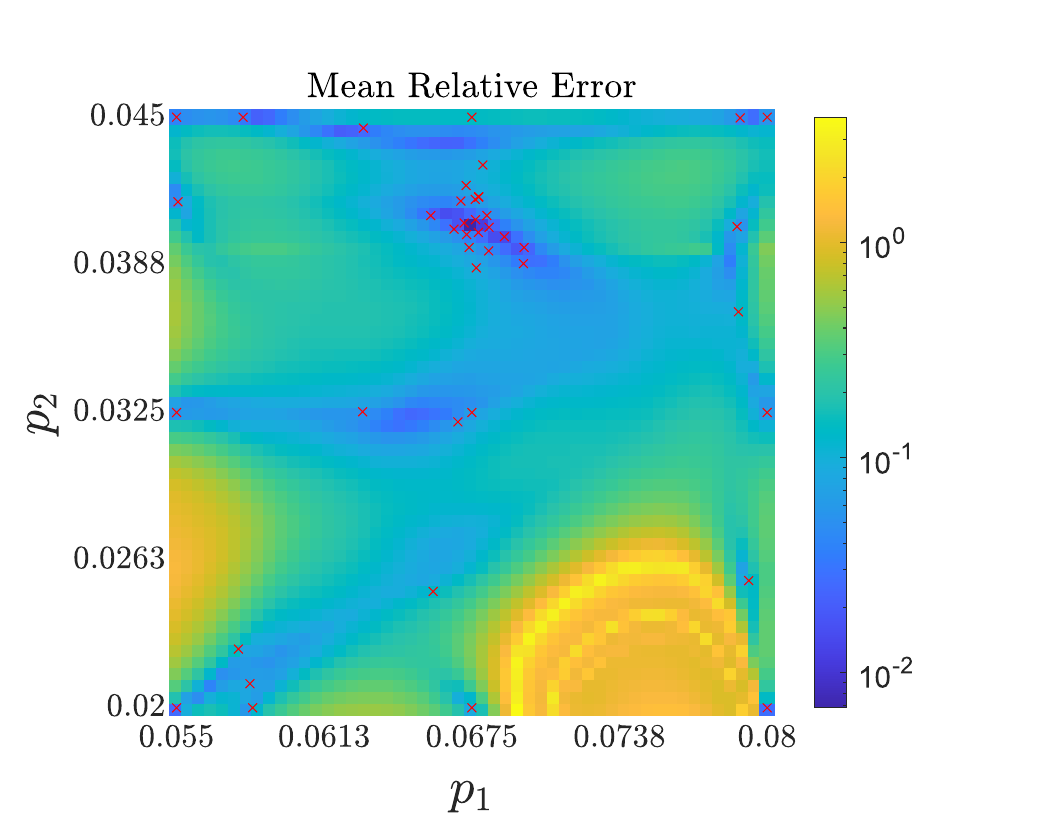}
            \label{fig:Run_7_Mean_Error}
        \end{adjustbox}
    \end{minipage}
    \caption{Final sample distribution of the Kelvin cell run and the corresponding mean error across the domain.
    \label{fig:KC2D_Final_3x3}
}
\end{figure}

Since this model shows a more complex dynamic behavior, the number of basis vectors to represent the model in the reduced basis must be chosen correspondingly (see Table \ref{tab:kelvincell_rom_parameters}). To start off, $r=30$ is chosen for the reduced local basis as this provided sufficient accuracy for the projected ROM. Initializing the algorithm with nine samples in a $3^2$ FFD yields globally defined reduction bases of size $r\approx 70$ after reprojection. Note that these reduced orders may increase when new samples are added and the concatenated matrix $\mathbf{V}_{\text{all}}$ grows in size.
\begin{table}[H]
    \centering
    \caption{Model order reduction parameters for the Kelvin cell example}
    \begin{tabular}{llll}
        \hline
        Parameter & Description & Value & Unit \\
        \hline
        $r$ & Reduced order model size & 30 & - \\
        $\varepsilon_{\text{IRKA}}$ & IRKA tolerance & $1 \times 10^{-6}$ & rad/s \\
        $N$ & Maximum IRKA iterations & 10 & - \\
        $f_{0, \text{IRKA}}$ & Initial expansion frequencies & linspace($-500$, $500$, $r$) & 1/s \\
        $\kappa_{SVD}$ & Retained information after re-projection & 0.9995 & - \\
        \hline
    \end{tabular}
    \label{tab:kelvincell_rom_parameters}
\end{table}
In Figure \ref{fig:KC2D_Final_3x3}, the optimization summary and the final sample distribution for the 2D Kelvin cell example for $f_\text{opt} = [300,400]$ Hz and nine starting samples are given.  The finer initial sampling distribution is chosen for two reasons: (1) The higher complexity of the model and the respective matrix entries simply require more samples in order to be able to find a proper fit for the matrix entries. (2) A too coarsely chosen initial grid can lead to "high-effort-medium-reward" iterations in which the found sample location does not justify the computational effort of global reduction, model training, and optimization. This phenomenon can be seen in the first iterations of the run from Figure \ref{fig:Beam_results}. Suppose the provided samples are entirely insufficient for a good approximation of the matrix entries, e.g., a linear model is fitted instead of a third-order polynomial. In that case, the quality of the transfer function approximation is correspondingly poor, and, more importantly, the relationship between the input parameters and the transfer function is not correctly modeled. Since this relationship is crucial for guiding the adaptive sampling in the regions of interest, the resulting samples are potentially far from the optimum. When these samples are set at the beginning of the algorithm, they still help refine the objective function by, for example, allowing a higher-order fit for the regression models. Therefore, the robustness of the algorithm is not impaired by these "low-value" samples that are set far away from the optimum, since they inherently help to explore the entire parameter domain. However, the global reduction and training of models, followed by a gradient-based optimization on an approximated objective, which might deviate significantly from the true objective, is a lot of computational overhead for obtaining a sample location that is highly sensitive towards the starting point of the gradient-based search and therefore just one random local minimum in the approximated objective surface.

\begin{figure}[H]
	\centering
		 {\includegraphics[width=0.99\textwidth]{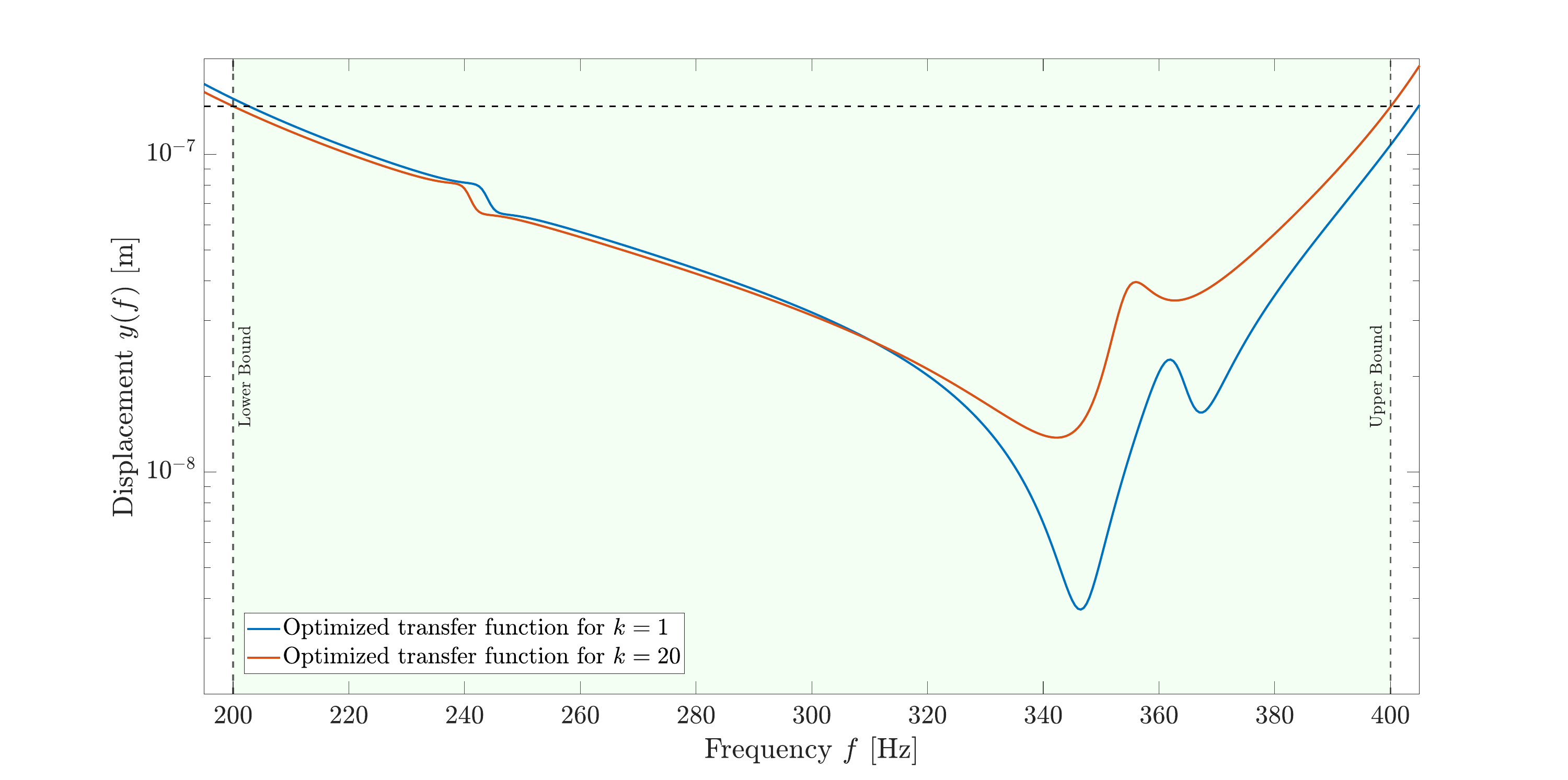}}
	\caption{Transfer function of the two optimal system configurations for $k=1$, and $k=20$ with $f_{\text{opt}} = [200,400]$ Hz.}
 \label{fig:TF_both_norms}
\end{figure}

Looking at the sample distribution and the corresponding mean error of the pROM, one can clearly see how the algorithm identifies the true optimum at $\boldsymbol{p}^*=[0.0673, 0.04]$ m and refines the model around that region. A total of 35 samples are set in addition to the nine starting samples. Roughly 13 samples are set far away from the optimum in the early stages, when the approximated objective surface does not resemble the true objective surface properly. With each subsequent iteration, the true objective shape becomes increasingly well represented, and the algorithm continues to place new samples in the vicinity of the optimum. The outcome of choosing a higher norm for the objective function shows in the final transfer function of the optimized system. While lower norms simply minimize the average or RMS (Root Mean Square) output, higher norms minimize the maximum output in the specified frequency range (see Figure \ref{fig:TF_both_norms}).

\subsection{Computational Performance}
\label{sec:computational_performance}
In this chapter, the computational performance of the adaptive sampling algorithm is briefly analyzed. For that, a total of 70 runs were conducted for the Kelvin cell example with the shown objective functions and different starting sample distributions. In all 70 runs, the provided framework successfully identified the true global optimum of the objective function. This robustness can be attributed to the "self-correcting" characteristic described in Section \Cref{subsec:KC_results}. The required times for the different subroutines and the number of iterations were measured and averaged. The computations were performed on an Intel Xeon(R) E5-2620 v3 processor with 2.4 GHz and MATLAB R2020a. The final algorithm can be utilized in two different ways: Firstly, as an adaptive sampling strategy in which the iterative optimization is used to determine new sample locations. In that case, the goal is to reduce the number of FOM generations required to achieve a locally accurate pROM. For that, the algorithm is compared to different non-adaptive sampling techniques like full grid and LHS. Another way of using the algorithm could be by focusing on the final optimization of the model while starting with a rather dense sampling distribution. For this, the benchmark is the "brute-force" optimization of the FOM or the ROM using finite differences. In the following, we will take a closer look at three approaches and their corresponding runtimes and sample counts. 

\subsubsection{Comparison to non-adaptive sampling}
A summary of the averaged runs for four example objectives is given in Table \ref{tab:averaged_runs}. The different runtimes for the respective objectives can firstly be attributed to the larger $f_{\text{opt}}$-range, which is discretized with 101 and 201 frequency points, respectively. This results in twice as many frequency points for which the output of the ROM needs to be determined. Interestingly, the choice of a higher $L_k$-norm also results in a significant increase in runtime, when compared to the number of samples added. This increase can predominantly be attributed to the gradient-based optimization subroutine, which shows an average 30\% higher runtime per added sample for the $L_{20}$-norm compared to the $L_1$-norm. The single optimization runs taking longer for a higher $L_k$-norm could be due to a multitude of reasons. Since the cost evaluation, as well as the analytical gradient calculation, contain the exact same operations but with different values for $k$, the amount of time required for these routines is roughly the same. The increased overall optimization time thus has to be attributable to the remaining subroutines that are involved in MATLAB's \lstinline{fmincon} function. These could include the finite difference approximation of the Hessian or the line search within the deployed interior-point algorithm. Since the objective is significantly more sensitive, the approximated Hessian contains large values. These large values paired with very small values result in smaller step sizes and can, in the worst case, introduce numerical instabilities.
\begin{table}
\centering
\caption{Sample and runtime overview for the different objectives}
\begin{tabular}{l|cccc}
\hline
\makecell{Objective} & \makecell{Number\\of runs} & \makecell{Initial\\samples $n_{IS}$} & \makecell{Added\\samples $n_{AS}$} & \makecell{Total\\runtime [s]} \\
\hline
 $f_{\text{opt}} = [300,\,400]$ Hz, $k=1$  & 20 & 9 & 33  & 3540 \\
 $f_{\text{opt}} = [300,\,400]$ Hz, $k=20$ & 20 & 9 & 28  & 3420 \\
 $f_{\text{opt}} = [200,\,400]$ Hz, $k=1$  & 20 & 9 & 34  & 3660\\
 $f_{\text{opt}} = [200,\,400]$ Hz, $k=20$ & 20 & 9 & 34  & 4890 \\
\hline
\end{tabular}
\label{tab:averaged_runs}
\end{table}
\begin{table}
\centering
\caption{Runtime overview for the different objectives for the non-adaptive sampling}
\begin{tabular}{l|ccc}
\hline
\makecell{Objective} & \makecell{Number\\of runs} & \makecell{Initial\\samples $n_{IS}$} &     \makecell{Total\\runtime [s]} \\
\hline
 $f_{\text{opt}} = [300,\,400]$ Hz, $k=1$  & 10 & 100  & 510\\
 $f_{\text{opt}} = [300,\,400]$ Hz, $k=20$ & 10 & 100  & 550\\
 $f_{\text{opt}} = [200,\,400]$ Hz, $k=1$  & 10 & 100  & 540\\
 $f_{\text{opt}} = [200,\,400]$ Hz, $k=20$ & 10 & 100  & 580\\
\hline
\end{tabular}
\label{tab:FFD_LHS_Benchmark}
\end{table}

\begin{figure}[H]
	\centering
		 {\includegraphics[width=0.6\textwidth, trim=2cm 8cm 3cm 8cm,clip]{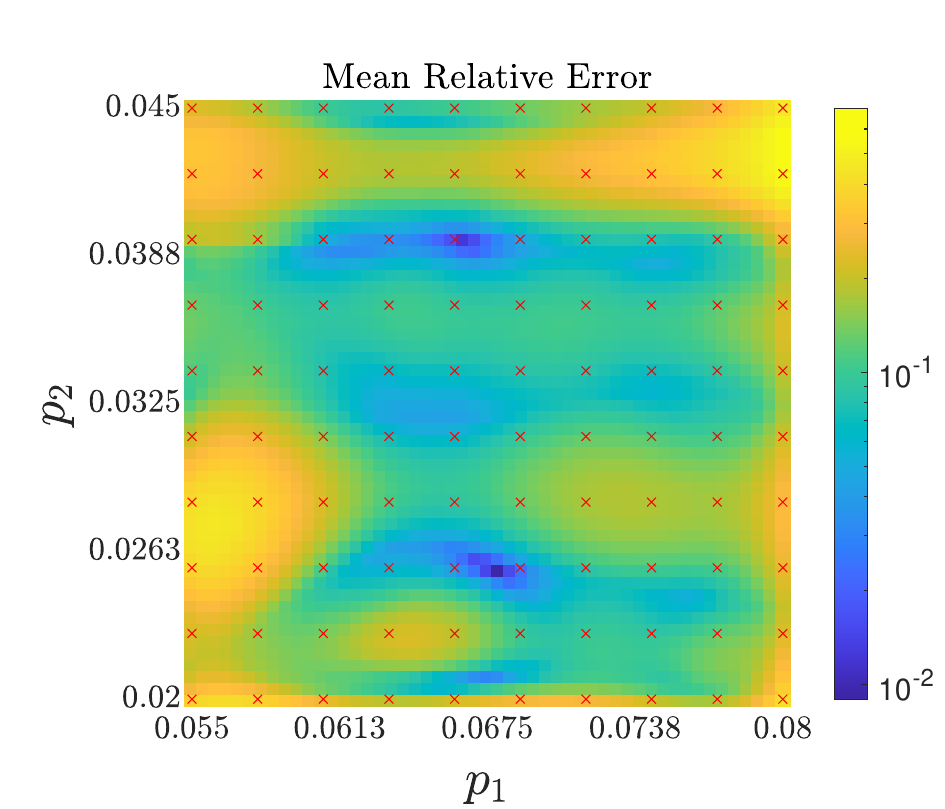}}
	\caption{Mean error of the pROM resulting from a non-adaptive $10^2$ FFD.}
 \label{fig:Mean_error_10x10}
\end{figure}

Table \ref{tab:FFD_LHS_Benchmark} shows averaged runtimes for the non-adaptive sampling routine, which in that case is a $10^2$ FFD. It can be seen that the adaptive procedure is not able to outperform the simple FFD sampling in terms of wall clock times. A dense initial sampling, which yields a sufficiently good approximation of the objective surface in the first iteration, significantly reduces the computation time since all FOMs can be generated using parallelization. The resulting accuracy of the pROM, however, is significantly lower than the local accuracy of the pROM resulting from the proposed adaptive scheme (see Fig. \ref{fig:Mean_error_10x10}), while still using more samples. While the adaptive scheme refines the model in the region around the optimum, the amount of total samples required to reach a certain accuracy is reduced. We can thus conclude that even though a reduction in wall clock time could not be achieved, the proposed framework consistently reaches higher local accuracies while using a lower number of samples.

\begin{figure}[H]
    \centering
    \begin{subfigure}[b]{0.45\linewidth}
        \centering
        \begin{subfigure}[b]{\linewidth} 
            \begin{tikzpicture}
            \begin{axis}[
                xbar,
                xmin=0,
                bar width=12pt,
                axis on top,
                bar shift=0pt,
                enlarge y limits=0.5,
                ytick={0,1},
                yticklabels={Sample adding, Optimi- zation}, 
                yticklabel style={anchor=east, align=center, text width=2cm}, 
                xlabel={Time per iteration [s]},
                height=5cm,
                width=\linewidth, 
                y=0.9cm,
            ]
               \addplot+[
                    fill=TUMBlue5,
                    draw=black,
                    error bars/.cd,
                    x dir=both,
                    x explicit,
                    error bar style={color=black},
                ]
                coordinates {(21.7,0) +- (0.9,0)};
                \addplot+[
                    fill=TUMBlue6,
                    draw=black,
                    error bars/.cd,
                    x dir=both,
                    x explicit,
                    error bar style={color=black},
                ]
                coordinates {(65.4,1) +- (13.7,0)};
            \end{axis}
            \end{tikzpicture}
            \label{fig:times_nested1}
        \end{subfigure}

        \begin{subfigure}[b]{\linewidth} 
            \centering
            \begin{tikzpicture}
            \begin{axis}[
                xbar,
                xmin=0,
                bar width=12pt,
                axis on top,
                bar shift=0pt,
                enlarge y limits=0.3,
                ytick={0,1,2},
                yticklabels={Reprojection, {Model training}, Local reduction}, %
                yticklabel style={anchor=east, align=center, text width=2cm}, 
                xlabel={Time per iteration [s]},
                height=7cm,
                width=\linewidth, 
                y=0.9cm,
            ]
            \addplot+[
                fill=TUMBlue3!80,
                draw=black,
                error bars/.cd,
                x dir=both,
                x explicit,
                error bar style={color=black},
            ]
                coordinates {(1.76,2) +- (0.21,0)};
            \addplot+[
                fill=TUMBlue2!70,
                draw=black,
                error bars/.cd,
                x dir=both,
                x explicit,
                error bar style={color=black},
            ]
                coordinates {(1.21,1) +- (0.23,0)};
            \addplot+[
                fill=TUMBlue1!70,
                draw=black,
                error bars/.cd,
                x dir=both,
                x explicit,
                error bar style={color=black},
            ]
                coordinates {(0.85,0) +- (0.14,0)};
            \end{axis}
            \end{tikzpicture}
            \label{fig:times_nested2}
        \end{subfigure}
        \caption{Averaged absolute runtimes per iteration.}
        \label{fig:Times_per_It}
    \end{subfigure}
    \hfill
    \begin{subfigure}[b]{0.45\linewidth} 
        \centering
        \raisebox{1cm}{ 
            \includegraphics[width=\linewidth]{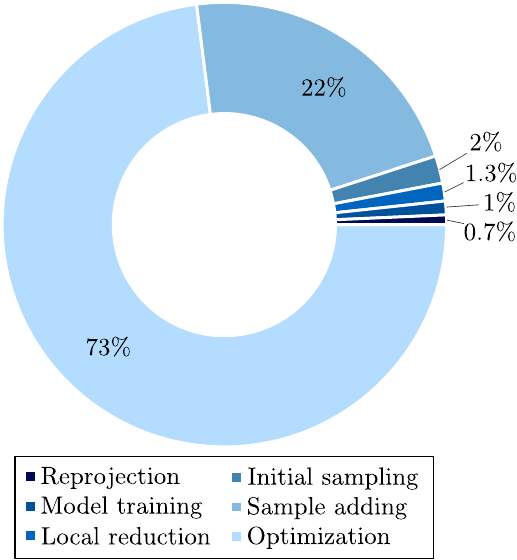} 
        }
        \caption{Averaged total runtime percentages.}
        \label{fig:donut_chart}
    \end{subfigure}
    \caption{Runtime analysis for the different subroutines.}
    \label{fig:Runtime_analysis}
\end{figure}

In order to better understand the increase in wall clock time, a more detailed analysis of the different subroutines deployed is given in the following. In Figure \ref{fig:Runtime_analysis}, the average runtime percentages and the absolute runtime of the different subroutines are portrayed. It can be seen that gradient-based optimization requires the most time (73 \%), followed by the sample-adding routine (22 \%). The initial sampling time accumulates to about 2 \%, while model training, local reduction, and reprojection account for approximately 1 \% of the total runtime each. Clearly, the gradient-based optimization is the bottleneck of the routine and determines the computation time. To investigate potential reasons for this, the optimization routine and the sample adding subroutines are now examined in more detail.

\textbf{Full Order Model Generation}
Generating a single FOM of the Kelvin Cell takes approximately 21 seconds. Since 12 threads are available for parallelization, a total of 12 FOMs can be generated at the same time. The total runtime required to generate the initial sampling $T_{IS}$ set can therefore be approximated by the formula \begin{equation}
    T_{IS}(n_{IS}) = t_S \times \left\lceil \frac{n_{IS}}{n_T} \right\rceil = 21\,\text{s} \times \left\lceil \frac{n_{IS}}{12} \right\rceil,
\end{equation}
with $t_{S}$ being the time required for the generation of one sample, $n_{IS}$ the number of initial samples, and $n_{T}$ the number of threads. For the starting distribution of a $3^2$ full factorial design, the parallelized model generation therefore takes about 21 seconds and thus makes up for roughly 2\% of the total runtime. A rather dense grid of $10 \times 10$ samples is generated in approximately 200 seconds. Since in the routine that extends the sample set, a single FOM is generated at the new sample location, the time for the sample adding routine $T_{AS}$ can be approximated by $T_{AS} = t_S \times n_{AS}$, with $n_{AS}$ being the number of added samples. This pattern is also confirmed when looking at the insignificant variance of the sample adding routine shown in \Cref{fig:Times_per_It}. 

\textbf{Gradient-Based Optimization}
When analyzing the computation time for the gradient-based optimization subroutine, the previously mentioned phenomenon of "high-effort-medium-reward" iterations (see Section \ref{subsec:KC_results}) needs to be addressed. Especially in the early stages of the algorithm, when a low number of FOMs are sampled, the objective surface highly deviates from the true response, exhibiting an almost noisy character (see Figure \ref{fig:Beam2D_Iteration01}). Running a gradient-based optimization on this surface will converge to local optima that are highly dependent on the starting point. Knowing now that the optimization routine is responsible for over 70\%, while local reduction and global reprojection barely account for 2\% of the computation time, the effort for the optimization routine cannot be justified for the early iterations. Considering that the times required to generate a denser grid are comparably low, it can also be argued that the early stages of the algorithm could be skipped entirely by starting with a dense 5x5 or 10x10 sampling.  It is important to note that these times correspond to the Kelvin cell example with $n_{DOF} \approx 10,800$. Potentially, there is a break-even point for even larger models, for which the time required for the adaptive sampling procedure falls below the time required to create a non-adaptive dense grid. To confirm this, a thorough analysis of the computational complexity of the adaptive sampling needs to be carried out.

Naturally, the time for the single optimization runs varies significantly since the starting points are chosen randomly, and thus, a different number of steps until convergence is required (see \Cref{fig:Runtime_analysis}). Furthermore, a certain variance remains due to the Thompson sampling, even when the same starting point is chosen. Notably, the single runs exhibit a growing trend in terms of runtime with advancing outer-loop iterations. This is due to the variable size of the ROM, which, as already mentioned, increases once more samples are added. A more thorough analysis of the separate steps inside the optimization routine shows that the gradient calculation, more precisely, the differentiation of the regression models for the matrix entries, can be identified as a reason for this. Towards the end of the adaptive sampling routine, the sample set contains around 40 FOM samples, which results in a global ROM size of $r \approx 150$. Since the upper triangular matrix is interpolated, the number of entries can be calculated by 
\begin{equation}
    n_{Mdl} = \sum_{i=1}^ri = 1+2+\dots+r = \frac{r(r+1)}{2},
\end{equation} 
which results in a quadratic increase of interpolated entries for growing ROM order $r$. An increase from $r=30$ to $r=150$ therefore results in 25 times as many matrix entry regression models that ought to be differentiated.

\subsubsection{Comparison to finite-difference optimization}
Following the idea of increasing the initial sample count in order to start off with a better approximation of the objective surface right away leads to the second approach that the algorithm can be utilized for. The provided framework can then be used for efficient optimization of the parametrized model. Deploying adjoint sensitivity analysis to calculate the analytical gradients of the objective functions by differentiating the system operators of the pROM efficiently cuts the average time spent on the optimization run in half, compared to the finite difference approximated gradients. This can be traced back to the count of objective function evaluations necessary to perform a single step in the gradient-based search, which is significantly lower for the analytical gradients. Therefore, the provided framework should be able to outperform both the FOM-, as well as the ROM-based optimization using finite difference approximations. Averaged computation times for this approach are provided in Table \ref{tab:ROM_FOM_Benchmark}.
\begin{table}[H]
\centering
\caption{Runtime overview for the finite difference optimization on ROM and FOM level}
\begin{tabular}{l|ccc}
\hline
\makecell{Objective} & \makecell{Number\\of runs} & \makecell{FOM: Total\\runtime [s]}&     \makecell{ROM: Total\\runtime [s]} \\
\hline
 $f_{\text{opt}} = [300,\,400]$ Hz, $k=1$  & 10 & 1197  & 1362 \\
 $f_{\text{opt}} = [300,\,400]$ Hz, $k=20$ & 10 & 2827  & 3689 \\
 $f_{\text{opt}} = [200,\,400]$ Hz, $k=1$  & 10 & 1180  & 1233 \\
 $f_{\text{opt}} = [200,\,400]$ Hz, $k=20$ & 10 & 2318  & 2685 \\
\hline
\end{tabular}
\label{tab:ROM_FOM_Benchmark}
\end{table}
When compared with Table \ref{tab:FFD_LHS_Benchmark}, it can be seen that for the optimization of the system, both the FOM and the ROM take a tremendous amount of time to converge. This is due to the many objective function evaluations necessary to conduct one optimization step. Averaging about seven function evaluations per iteration, this approach results in around 110 FOMs, which need to be generated and solved before the optimum is found. For the ROM-based optimizations, the reduction introduces further computational effort, which even results in an increase in computation time. This shows that for the optimization, a non-parametric ROM is not of good use unless a very large frequency range is involved in the optimization objective. While the proposed algorithm is not able to outperform a simple non-adaptive sampling, the included optimization framework leveraging the pROM and analytical gradients achieves a significant speedup compared to the finite-difference-based optimization of the FOM and the ROM.

\section{Discussion \& Conclusion}
\label{sec:conclusions}
This work proposed an optimization-driven adaptive sampling approach for projection-based reduced order models of dynamical systems. It combines matrix-interpolatory MOR, Bayesian regression, Thompson-sampling and adjoint sensitivity analysis to guide new observations into regions of high fitness, reducing FOM evaluations as well as the ROMs error in regions of interest. Exploration and exploitation are balanced by accounting for the uncertainties of the obtained regression models. Numerical results of the Timoshenko beam and the Kelvin cell structural topology demonstrate the framework's robustness in identifying the correct optimum and a consistent reduction in required sampling points compared to a deterministic, non-adaptive sampling scheme. By incorporating the parametrized model's fitness into the acquisition criterion, higher local accuracies of the pROM can be achieved with a smaller number of samples. Nevertheless, the computational overhead introduced by subroutines like the sparse Bayesian regression and the gradient-descent algorithm prevents an overall speed-up in wall clock time. Potential topics for further investigation therefore include targeted hyperparameter tuning and subroutine optimization. Additionally, a systematic analysis of the computational complexity of the algorithm can help identify problem regimes in which the reduction of full-order evaluation outweighs the additional cost introduced by the adaptive procedure. Furthermore, revisiting the sparsification strategy of the regression models can help unify the derivative evaluation, which corresponds to one of the innermost loops of the adaptive sampling algorithm. Finally, a direct comparison to "classical" Bayesian optimization techniques, including a dedicated acquisition function that is maximized, can yield further insights regarding sample efficiency, computational cost, and approximation quality. 

\section*{CRediT authorship contribution statement}
\textbf{Marcel Warzecha}  Writing - original draft, Methodology, Software, Validation, Visualization. \textbf{Sebastian Resch-Schopper:} Conceptualization, Supervision, Writing - original draft. \textbf{Gerhard Müller:} Supervision, Writing - review \& editing.

\addcontentsline{toc}{section}{References}
\bibliographystyle{plainurl}
\bibliography{bibtex/references}

\begin{appendices}

\section{Gradient of the Objective Function}
\label{sec:appendix_gradient}

We start the derivation of the gradient of the objective function with the concatenated system of subproblems
\begin{equation}
\tilde{\mathbf{K}}_{\text{tot}} \cdot \mathbf{x}_{\text{tot}} = \mathbf{f}_{\text{tot}},
\label{eqn:tot_sys}
\end{equation}
which is also shown in \Cref{eqn:tot_sys}. On this total system of equations, the adjoint problem can be formulated analogously to the single problems as derived in \Cref{eq:SA_Gradient2}. Note that this representation is only used for the intuitive derivation of the gradient expression. For efficiency reasons, all subproblems are handled separately in the code framework. The gradient of the response is then given by
\begin{equation}
     \frac{\mathrm{d}R}{\mathrm{d}\mathbf{p}} = \frac{\partial{R}}{\partial{\mathbf{p}}} + \left(\frac{\partial{R}}{\partial{\mathbf{x}_{\text{tot}}}}\right)^\top \frac{\partial{\mathbf{x}_{\text{tot}}}}{\partial{\mathbf{p}}},
     \label{eqn:Gradient}
\end{equation}
with 
\begin{equation}
\frac{\partial\mathbf{x}_{\text{tot}}}{\partial\mathbf{p}} = \tilde{\mathbf{K}}_{\text{tot}}^{-1}\left(\frac{\mathrm{d}\mathbf{f}_{\text{tot}}}{\mathrm{d}\mathbf{p}} - \frac{\mathrm{d}\tilde{\mathbf{K}}_{\text{tot}}}{\mathrm{d}\mathbf{p}} \mathbf{x}_{tot}\right).
\end{equation}
In the examples considered in this work, geometric parameters are used, so the applied force $\mathbf{f}$ is parameter-independent. Furthermore, the response does not depend on the parameters explicitly, so $\frac{\partial{R}}{\partial{p}} = 0$. \Cref{eqn:Gradient} can then be written as
\begin{equation}
 \frac{\mathrm{d}R}{\mathrm{d}\mathbf{p}} = \frac{\partial{R}}{\partial{\mathbf{x}_{\text{tot}}}} \tilde{\mathbf{K}}_{\text{tot}}^{-1}\left( - \frac{\mathrm{d}\tilde{\mathbf{K}}_{\text{tot}}}{\mathrm{d}\mathbf{p}} \mathbf{x}_{\text{tot}}\right). 
\end{equation}
While the inverse of the total dynamic stiffness matrix $\tilde{\mathbf{K}}_{\text{tot}}^{-1}$ and the concatenated state vectors of all subproblems $\mathbf{x}_{\text{tot}}$ are directly given by the total system, the derivative of the system matrices w.r.t. the parameters $\frac{\mathrm{d}\tilde{\mathbf{K}}_{\text{tot}}}{\mathrm{d}\mathbf{p}}$ will simply be solved by differentiating the regression models. Choosing Thompson sampling instead of a separate acquisition ensures that analytic differentiation of the Bayesian regression models remains feasible. \\
What remains to be derived is the expression for $\frac{\partial{R}}{\partial{\mathbf{x}_{\text{tot}}}}$.
Differentiating the response with respect to the total solution vector 
\begin{equation}
\frac{\partial}{\partial \mathbf{x_{\text{tot}}}}\left[ \left( \sum_{i}^{n} |\mathbf{g}_{i} \cdot \mathbf{x}_{i}|^k\right)^{\frac{1}{k}} \right],
\end{equation}
can be approached by using the chain rule and differentiating and building the sum over the single outputs $|y_{i}| =|\mathbf{g}_{i} \cdot \mathbf{x}_i|$:
\begin{equation}
\frac{\partial{R}}{\partial{\mathbf{x}}_{\text{tot}}} = \sum_{i}^{n} \left[\frac{\partial{R}}{\partial|{y}_{i}|} \cdot \frac{\partial{|y_i|}}{\partial{y_i}} \cdot \frac{\partial{{y}_{i}}}{\partial{\mathbf{x}}_{\text{tot}}}\right].
\end{equation}
Thereby, 
\begin{equation}
    \frac{\partial{R}}{\partial|{y}_{i}|} = \left( \sum_{j=1}^{n} |y_j|^k \right)^{\frac{1}{k} - 1} \cdot |y_i|^{k-1} ,
\end{equation}
and
\begin{equation}
     \frac{\partial y_i}{\partial \mathbf{x}_{\text{tot}}} = \begin{bmatrix} \mathbf{0} & \cdots & \mathbf{g}_i & \cdots & \mathbf{0} \end{bmatrix}^\top.
\end{equation}
Since the output $y_i$ is a complex variable, the Wirtinger derivative is required to compute $\frac{\partial{|y_i|}}{\partial{y_i}}$, which results in 
\begin{equation}
    \frac{\partial{|y_i|}}{\partial{y_i}} = \Re \left[ \frac{\bar{y_i}}{|y_i|} \right],
    \label{eqn:real_valued_abs_derivative}
\end{equation}
according to \Cref{eq:wirt_2Re} and (\ref{eq:Wirtinger}). The analytic expression for the derivative of the response w.r.t. the solution vector thus results in
\begin{equation}
    \frac{\partial{R}}{\partial{\mathbf{x}}_{\text{tot}}} = \sum_{i}^{n} \left[\left( \sum_{j=1}^{n} |y_j|^k \right)^{\frac{1}{k} - 1} \cdot |y_i|^{k-1} \cdot \underbrace{\frac{1}{2}\frac{\bar{y_{i}}}{  |y_{i}|}}_{\text{Wirtinger Derivative}} \cdot \begin{bmatrix} \mathbf{0} & \cdots & \mathbf{g}_i & \cdots & \mathbf{0} \end{bmatrix}^\top\right],
    \label{eqn:dRdx_damped}
\end{equation}
yielding the final gradient expression
\begin{equation}
\frac{\mathrm{d}R}{\mathrm{d}\mathbf{p}} = \Re \Bigg(\sum_{i}^{n} \left[\left( \sum_{j=1}^{n} |y_j|^k \right)^{\frac{1}{k} - 1} \cdot |y_i|^{k-1} \cdot \frac{\bar{y_{i}}}{  |y_{i}|} \cdot \begin{bmatrix} \mathbf{0}  \cdots  \mathbf{g}_i  \cdots  \mathbf{0} \end{bmatrix}^\top\right] \cdot \tilde{\mathbf{K}}_{\text{tot}}^{-1}\left( - \frac{\mathrm{d}\tilde{\mathbf{K}}_{\text{tot}}}{\mathrm{d}\mathbf{p}} \mathbf{x}_{\text{tot}}\right)\Bigg).
\label{eqn_damped_gradient}
\end{equation}

\end{appendices}

\end{document}